\documentclass[conference, 10pt]{IEEEtran}
\IEEEoverridecommandlockouts
\usepackage{cite}
\usepackage{amsmath,amssymb,amsfonts}
\usepackage{algorithmic}
\usepackage{graphicx}
\usepackage{textcomp}
\usepackage{xcolor}
\usepackage{subcaption}
\usepackage{tikz}
\usepackage{tabularx}
\usepackage{multirow}
\usepackage{hhline}
\usepackage{makecell}
\usepackage{moresize}
\usepackage{varwidth}

\def\BibTeX{{\rm B\kern-.05em{\sc i\kern-.025em b}\kern-.08em
    T\kern-.1667em\lower.7ex\hbox{E}\kern-.125emX}}

\newcolumntype{Y}{>{\centering\arraybackslash}X}
\newlength{\Oldarrayrulewidth}

\begin{document}


\newcommand{\xxx}[1]{{\color{red}{[\textbf{XXX}: #1]}}}
\newcommand{\todo}[1]{{\color{red}{[\textbf{TODO}: #1]}}}
\newcommand{\reword}[1]{{\color{cyan}{[\textbf{REWORD}: #1]}}}
\newcommand*\circled[1]{\tikz[baseline=(char.base)]{
            \node[shape=circle,draw,inner sep=2pt] (char) {#1};}}
            
\newcommand{\changed}[1]{#1}

\newcommand{\RMone}{DRM1}
\newcommand{\RMtwo}{DRM2}
\newcommand{\RMthree}{DRM3}
\newcommand{\remoteshard}{sparse shard}
\newcommand{\SCLarge}{SC-Large}
\newcommand{\SCSmall}{SC-Small}
\newcommand{\fbthrift}{\emph{Thrift}}
\newcommand{\facebook}{Facebook}
\newcommand{\dlrm}{deep recommendation system}
\title{Understanding Capacity-Driven Scale-Out Neural Recommendation Inference\\
}

\author{
    \IEEEauthorblockN{
        Michael Lui\IEEEauthorrefmark{2}\IEEEauthorrefmark{1},
        Yavuz Yetim\IEEEauthorrefmark{1},
        \"Ozg\"ur \"Ozkan\IEEEauthorrefmark{1},
        Zhuoran Zhao\IEEEauthorrefmark{1},
        Shin-Yeh Tsai\IEEEauthorrefmark{1},\\
        Carole-Jean Wu\IEEEauthorrefmark{1},
        Mark Hempstead\IEEEauthorrefmark{1}\IEEEauthorrefmark{2}
    }
    \IEEEauthorblockA{
        \IEEEauthorrefmark{1}Facebook,
        \IEEEauthorrefmark{2}Drexel University,
        \IEEEauthorrefmark{2}Tufts University
    }
}

\maketitle

\begin{abstract}
Deep learning recommendation models have grown to the terabyte scale.
Traditional serving schemes--that load entire models to a single server--are unable to support this scale.
One approach to support this scale is with distributed serving, or distributed inference, which divides the memory requirements of a single large model across multiple servers.

This work is a first-step for the systems research community to develop novel model-serving solutions, given the huge system design space.
Large-scale deep recommender systems are a novel workload and vital to study, as they consume up to 79\% of all inference cycles in the data center.
To that end, this work describes and characterizes scale-out deep learning recommendation inference using data-center serving infrastructure.
This work specifically explores latency-bounded inference systems, compared to the throughput-oriented training systems of other recent works. 
We find that the latency and compute overheads of distributed inference are largely a result of a model's static embedding table distribution and sparsity of input inference requests.
We further evaluate three embedding table mapping strategies of three DLRM-like models and specify challenging design trade-offs in terms of end-to-end latency, compute overhead, and resource efficiency.
Overall, we observe only a marginal latency overhead when the data-center scale recommendation models are served with the distributed inference manner--P99 latency is increased by only 1\% in the best case configuration.
The latency overheads are largely a result of the commodity infrastructure used and the sparsity of embedding tables.
Even more encouragingly, we also show how distributed inference can account for efficiency improvements in data-center scale recommendation serving.

\end{abstract}


\section{Introduction}

\begin{figure}[t]
	\centering
	\includegraphics[width=\linewidth]{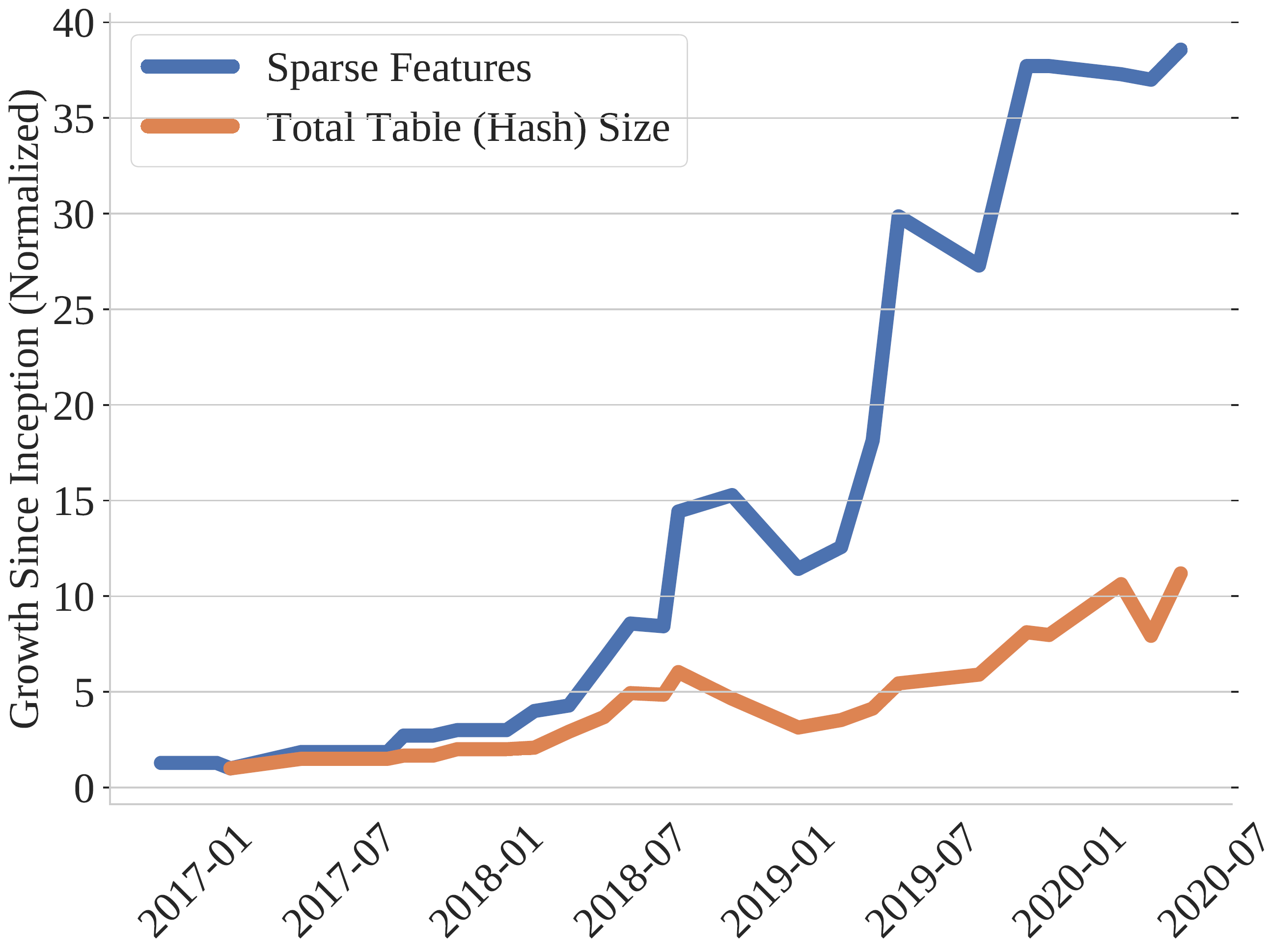}
	\caption{Historical model growth of significant production recommendation model. Both number of features and embeddings have grown an order of magnitude in only three years.}
	\label{fig:model_growth}
\end{figure}

Deep learning recommendation models play an important role in providing high-quality internet services, and has recently started receiving attention from the systems community~\cite{fb_rec_hpca,guptadeeprecsys_isca20,fb_dlrm, fb_bandana, fb_recnmp, hwang2020centaur,kwon2020tensor,hsia:iiswc20}. 
Recommendation workloads have been shown to consume up to 79\% of all AI inference cycles at Facebook's data-centers~\cite{fb_rec_hpca} in 2019 and have been added to the MLPerf benchmarking effort~\cite{mlperf-inference,mlperf-training,mlperf-reco-advisory}.
The significance of this data-center workload warrants increased attention to its performance, throughput, and scalability. 
A major challenge facing this distinct class of deep learning workload is their growing size.
Figure~\ref{fig:model_growth} shows the rate of recommendation model growth at
\facebook. Further models at the 1-10TB scale have also been deployed at Baidu and Google~\cite{baidu_dist_inf, baidu_aibox, deepretrieval}.
\changed{This growth is spurred by the desire for increased accuracy. 
Increasing model size, as number of parameters, is well known to improve generalization and accuracy and is necessary to achieve optimal results given larger training data sets~\cite{gshard, scaling_dl, scaling_dl_beyond, scaling_laws_lm, understanding_dl_generalization, Geiger_2020}.
This is true across of a variety of deep learning applications, including language models with large embedding tables--similar to \dlrm{}s.
This work does not explore the accuracy effects of increasing model size and instead focuses on \textit{system and performance} implications of supporting the massive models used in production \dlrm{}s.}

Deep recommendation model size, and by proxy memory capacity, is dominated by \textit{embedding tables}, which represent learned sparse parameters.
Each sparse input is hashed to one or more indices in its corresponding embedding table, where each index maps to a vector.
Indexed embedding vectors are then combined in a pooling operation, such as a vector summation.
Increasing hash bucket size and the number of embedding tables increase the information captured by the embeddings and thus is a straightforward method to improve the accuracy of recommendation models, but is constrained by the commensurate increase in memory footprint~\cite{fb_compositional_embeddings}.
Reducing hash bucket size will constrain model size, and so must be performed with care to maintain model accuracy.

\begin{figure*}[t]
	\centering
	\begin{subfigure}[c]{.45\linewidth}
		\includegraphics[width=\linewidth]{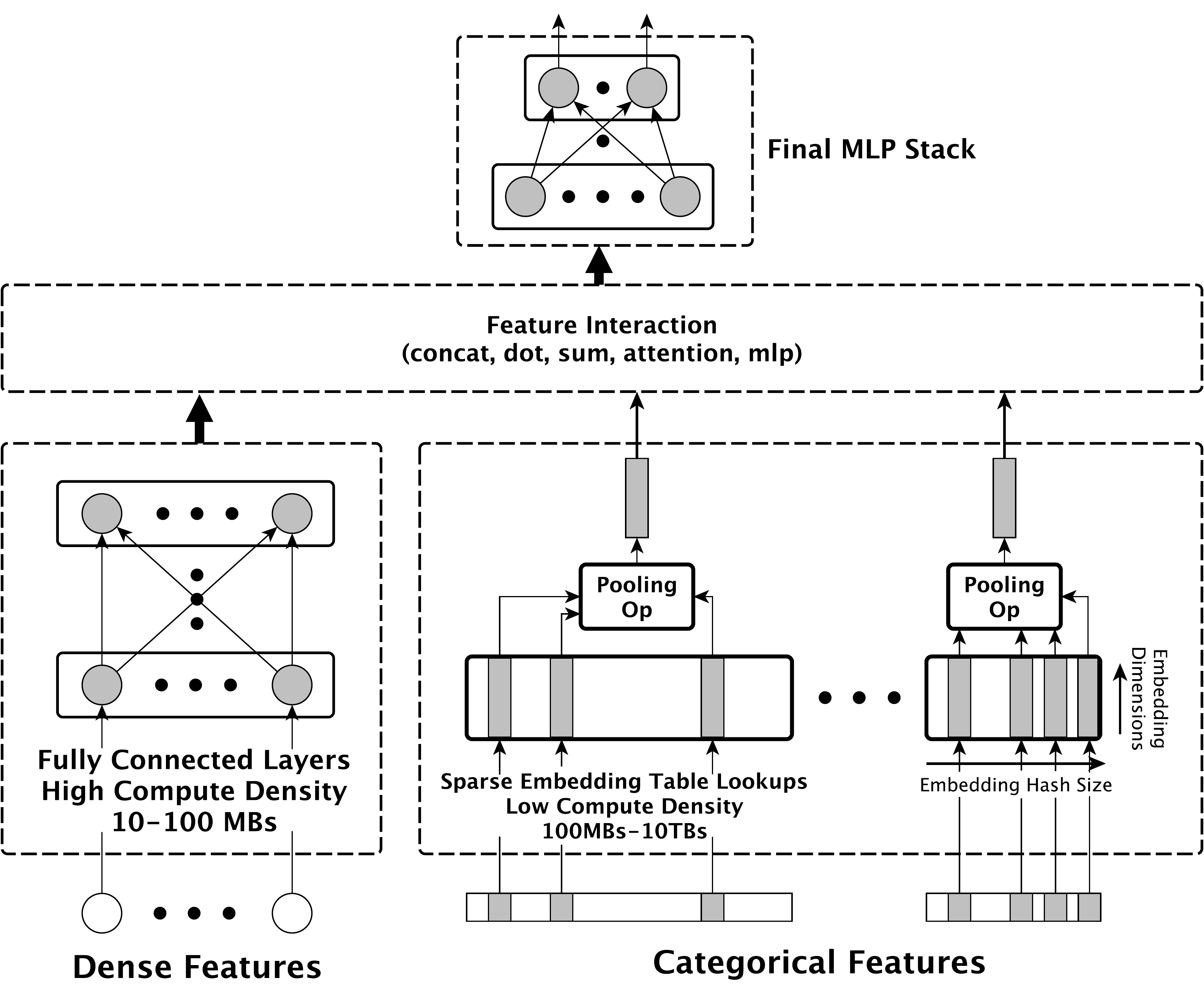}
		\caption{A simplified recommendation ranking model. Sparse features transformed into dense values via embedding table lookups.}
		\label{fig:sparsenn}
	\end{subfigure}
	\qquad
	\begin{subfigure}[c]{.45\linewidth}
		\includegraphics[width=\linewidth]{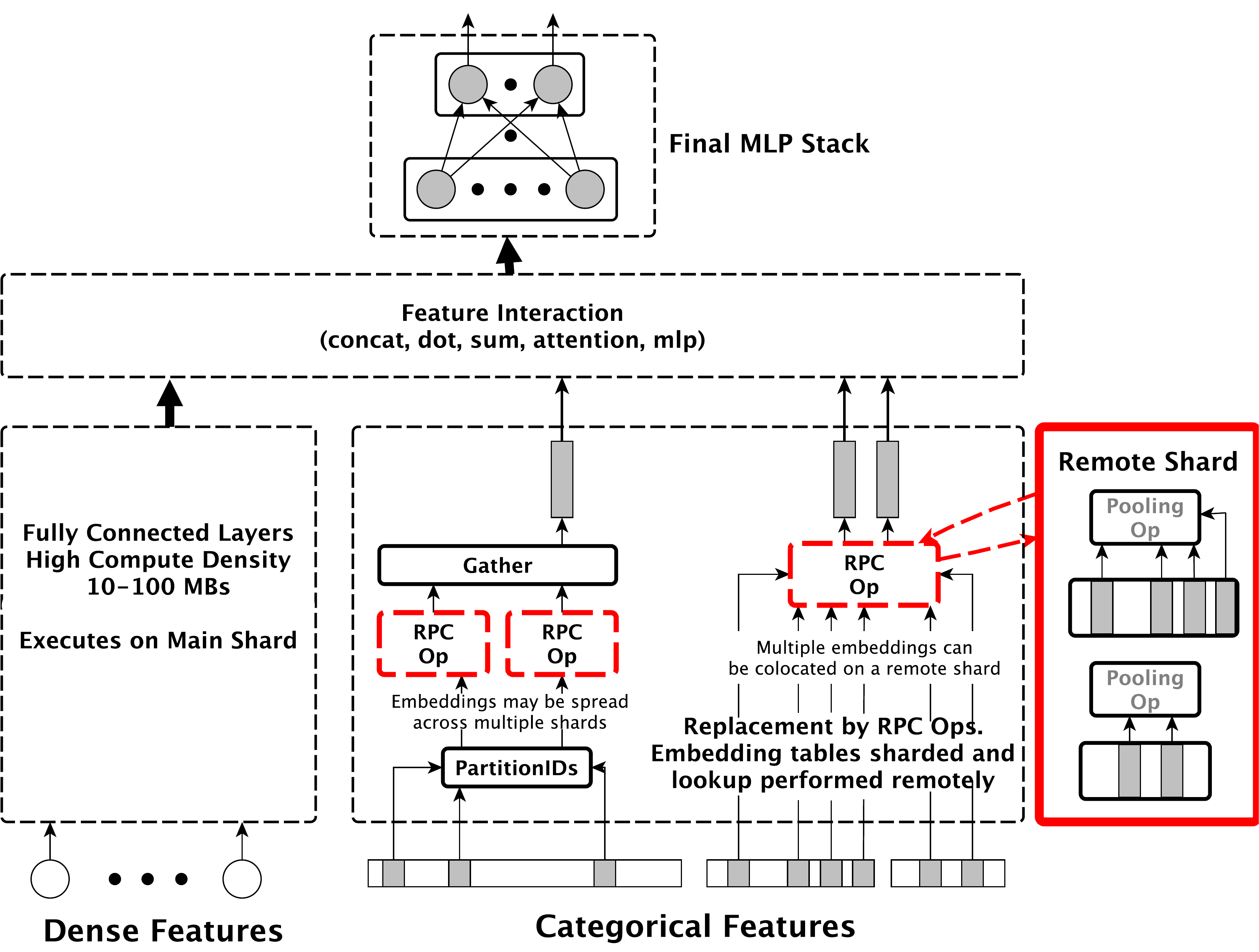}
		\caption{The recommendation model sharded into parts due to capacity limitations. Sparse lookup operators have been replaced with asynchronous RPC ops. Each remote procedure call may query one, multiple, or fractional tables.}
		\label{fig:sparsenn_shards}
	\end{subfigure}
	
	\caption{The recommendation network in \ref{fig:sparsenn} is partitioned, or sharded, to run on multiple servers, in \ref{fig:sparsenn_shards}. Heuristically, the sparse operators are placed on \remoteshard{}s because 1) they consume the majority of memory resources and 2) are independent of the dense operators and stateless. Some tables are split across multiple shards because they are larger than system memory can support. }
	\label{fig:dist_sparsenn}
\end{figure*}

As memory requirements for large recommendation models surpass the memory capacity of a single server, solutions are needed to either constrain model size or increase effective memory capacity.
Model compression techniques can be used to constrain model size but can degrade model accuracy~\cite{deep_compression_iclr}.
DRAM capacity can be increased, but this is not scalable beyond the single-digit terabyte range. 
Even then, larger server chassis are required to support large memory capacity. 
On demand paging of the model from higher capacity storage is another solution, but this requires fast solid-state drives (SSD) to meet latency constraints. 

Additional hardware requirements are undesirable in the data-center due to the added complexity of managing heterogeneous resources.
For example, clusters with specialized configurations cannot easily expand resources during periods of high activity or efficiently shrink resources during periods of low activity.
This is particularly true of workloads affected by diurnal traffic patterns through the day~\cite{kim2018hpca}.
In contrast to the aforementioned approaches, a distributed serving paradigm can be implemented on existing infrastructures by splitting the model into distinct and independent sub-networks to run on common CPU platforms.
Of course, this approach is not without tradeoffs.
Added network latency, network load, and the requirement of additional compute nodes are incurred as a result.
However in a large homogeneous data-center, it is easier to deploy and scale to these resources compared to custom, unconventional hardware platforms.

As such, we provide the research community in-depth descriptions and characterizations of distributed inference for deep learning recommender systems.
It is a first approach, scalable solution to the challenge of growing model sizes and presents a baseline system for further optimization.
Notably, it is distinct from recent throughput-oriented \emph{training} systems that do not have the same latency constraints found in inference serving~\cite{baidu_dist_inf, gshard}.
The implementation is characterized on real inputs and real, scaled-down deep learning recommendation models.
Three model parallelization strategies are explored within this serving paradigm.
This challenging, novel workload presents new opportunities for the systems research community in the data-center scale, machine learning domain.
The contributions of this work are as follows:
\begin{itemize}
	\item To our knowledge, this is the first work describing a distributed inference serving infrastructure for at-scale neural recommendation inference.
	\item We present an in-depth characterization and breakdown of distributed inference's impact on end-to-end recommendation serving latency, tail latency, and operator compute overhead.  We further investigate the impact of embedding table placement using model sharding and place our findings in the context of the data-center serving environment.
	\item Finally, we design a cross-layer, distributed instrumentation framework for performance debugging and optimization analysis to quantify the performance overhead from remote procedure call (RPC) services and the machine learning framework.
\end{itemize}

The paper is organized as follows:
Section~\ref{sec:background} provides a brief introduction to deep learning recommendation.
Section~\ref{sec:dist_inf} describes how distributed inference is applied and implemented. Special attention is given to the sharding strategies used to generate distributed models.
Section~\ref{sec:tracing_framework} describes the custom tracing framework used to collect workload measurements.
Section~\ref{sec:methodology} provides details about the models, platforms, and inputs used in our characterization.
Sections~\ref{sec:results} and \ref{sec:data_center} present our characterization findings and place them in context of a data-center serving environment. We identify guidances and key takeaways for system designers.
Section~\ref{sec:related_work} provides a discussion of related works.
Finally, Sections~\ref{sec:academic_relevance} and \ref{sec:conclusion} offers our concluding thoughts.
\section{Recommendation Inference At-Scale}
\label{sec:background}

Recommendation is the task of recommending, or suggesting, product or content items from a set that are of interest to a user.
For example, a service may recommend new video clips, movies, or music based on a user's explicitly \textit{liked} content and implicitly consumed content.
Accuracy of the model is an abstract measure of a user's interest and satisfaction in the recommended results.
Traditionally, neighborhood based techniques like matrix factorization have been used to good effect by providing recommendation based on similarity to other users or similarity of preferred items~\cite{mf_rec, nh_rec}.
However, more recent recommender systems have used deep neural networks to combine a variety of dense and sparse input features into a more generalized predictive model~\cite{fb_dlrm, fb_rec_hpca, wide_n_deep, youtube_rec, din_rec, deepfm}.
An example of a dense feature is a user's age; an example of a sparse, or categorical, feature is web pages the user likes, and the output of the model are rankings of the candidate item inputs.
Figure~\ref{fig:sparsenn} shows a simplified overview of this deep learning recommendation model architecture.

Today, recommendation \emph{inference} is performed on the CPU, compared to the heterogeneous systems popular with other deep learning workloads~\cite{fb_rec_hpca, guptadeeprecsys_isca20}.
 This is because, as compared to GPUs and other AI accelerators leveraged in other inference tasks, the (1) sparsity in recommendation models, (2) evolving nature of recommendation algorithms, and (3) latency-bounded throughput constraints make it challenging for recommendation inference be served on AI accelerators efficiently at-scale.
Throughput, or queries per second (QPS), is a paramount target for inference, but just as important are latency constraints.
In order to provide a satisfactory user experience, recommendation results are expected within a timed window.
This strict latency constraint defines the service-level agreement (SLA)~\cite{fb_rec_hpca}.
If SLA targets cannot be satisfied, the inference request is dropped in favor of a potentially lower quality recommendation result, which could worsen user experience.
To maximize throughput within the respective SLA constraints, various techniques are applied, such as batch sizing and model hyperparameter tuning.

\subsubsection{Sparse Feature Operation}
Dense features are processed with fully-connected (FC) layers, while sparse features, shown in Figure~\ref{fig:sparsenn}, go through a transformation before further feature interaction.
The sparse inputs are transformed into a list of access IDs, or hash indices, which index into an \textit{embedding table}.
The size of the embedding table--the number of buckets and the vector dimension--is a tunable hyperparameter.
Usually, there is one embedding table per sparse feature, but it is possible for features to share tables to save memory resources.
For example, in a $N\times M$ table, with embedding dimension $M$, a sparse feature input with $K$ indices will produce $K$ \textit{embedding vectors} of length $M$.
A pooling operation, such as summation or concatenation, will collapse the matrix along the first dimension to produce a $1\times M$ or $1 \times (M K)$ result for use in the feature interaction.
In the Caffe2 framework, used in this work, the embedding table operator is called \textit{SparseLengthsSum}, or SLS.
It is typical to refer to the \textit{family} of related operators also as SLS.
Because of the sheer number of possible inputs, the embedding tables are constrained in size and hash collisions may occur.
For example, one FP32 embedding table with a dimension of 32, for 3-billion unique users, will consume over 347GB.
The table size would need to be reduced to tractably use on a single server.

\subsubsection{Substantial Model Growth and Large Embedding Tables}
To improve model accuracy and capitalize on the rich, latent information in sparse features, deep learning recommendation models have exploded in size as shown in Figure~\ref{fig:model_growth} and recent works~\cite{deepretrieval, baidu_dist_inf}.
The embedding tables dominate the size of recommendation models and are responsible for the significant growth in model size.
As rapid as this growth appears, it is still constrained by the commensurate increase in the memory capacity requirement.
In an effort to further improve the accuracy of recommendation models, \textit{the total capacity of embedding tables has grown larger than can be supported on a single server node, motivating the need for distributed serving paradigms}.

 \begin{figure*}[t]
	\centering
	\includegraphics[width=\textwidth]{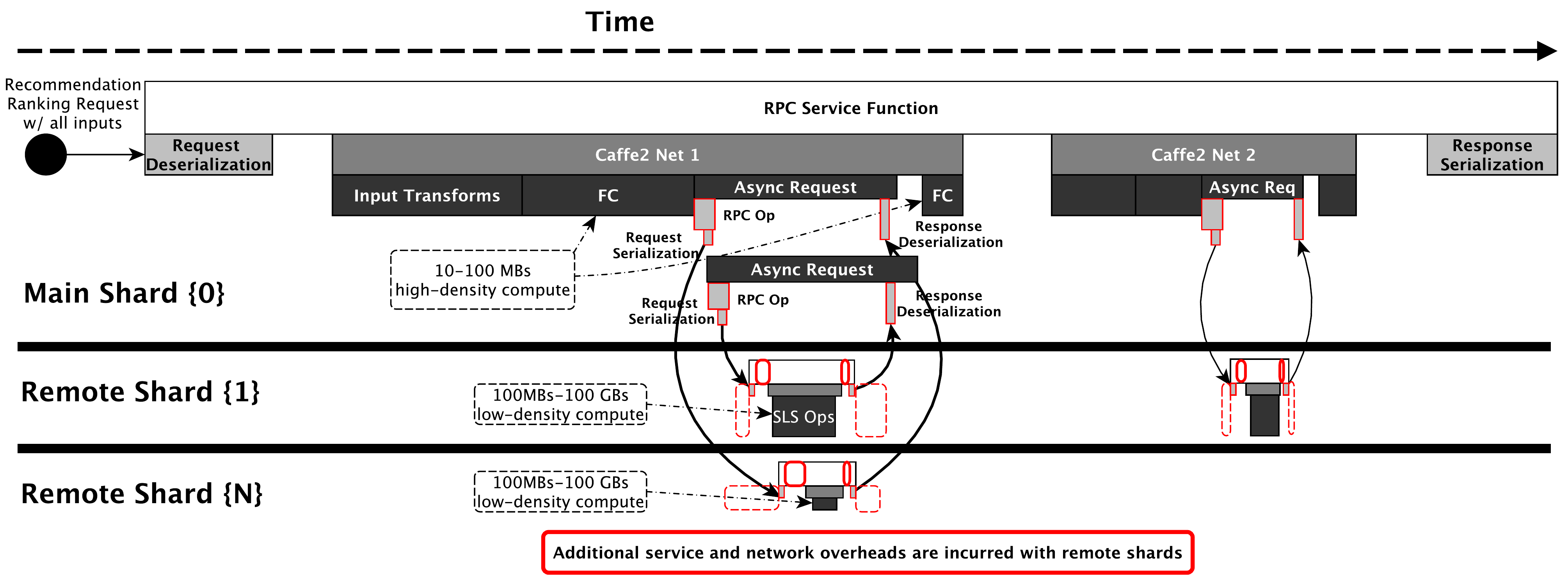}
	\caption{Example Trace of Distributed Inference. Layers are not to scale and do not represent the full set. All inference requests are forwarded to the main shard, which then invokes \remoteshard{}s when an RPC operator is encountered. The asynchronous nature enables an additional level of parallelism.}
	\label{fig:dist_trace}
\end{figure*} 

\section{Distributed Inference}
\label{sec:dist_inf}

This is the first work to describe distributed, neural network inference in detail for the unique goals and characteristics of \textit{\dlrm{}s}.
Because this initial implementation targets deployability and scalability over performance and efficiency, many optimization opportunities exist across the systems design space, including algorithms, software, microarchitecture, system configuration, and task mapping.
Thus, in this section, we provide an overview of the general system design and structure for the research community.
At minimum, a distributed inference solution for neural recommendation must enable (1) a greater variety and a larger number of sparse features and (2) larger embedding tables, i.e. larger embedding dimensions, less hash collisions, and/or less compression.

\subsection{Distributed Model Execution}
\label{sec:dist_inf_serving}

Consider that a deep learning model can be represented by a directed control and dataflow graph.
A distributed model simply partitions a neural network into subnets, where each subnet can operate independently.
Treating each subnet independently provides desirable flexibility in deployment, since all the infrastructure for serving the model already exists.
An example of such partitioning is shown in Figure~\ref{fig:sparsenn_shards}.
This is traditionally referred to as model or layer parallelism.

\subsubsection{Model Sharding}
\label{sec:model_sharding}
We call each independent, partitioned subnet a \textit{shard}, and the process of creating shards, \textit{sharding}.
Sharding occurs both before and after training.
Before training, \textit{parameter server} shards are generated to hold model parameters for distributed training.
After training, during model publishing, parameters are \textit{resharded} and serialized from parameter servers to the respective inference shard based on a prior partitioning phase.
At this stage, other model transformations, e.g. quantization, are executed and all training metadata is already available for sharding decisions.
Resharding directly after training also avoids the extra storage, compute, and complexity needed to reload and reshard terabyte-scale models.
In Figure~\ref{fig:sparsenn_shards}, the \textit{main shard} performs all the dense layers, and any partitioned subnets are replaced by custom remote-procedure-call (RPC) operators that call \textit{remote shards}.

Only the sparse operators, like SLS, and their embedding tables are placed on remote shards.
This heuristic directly addresses the memory constraint imposed by embedding tables and retains compute density on the main shard.
Thus, remote shards are also termed \remoteshard{}s.
Figure~\ref{fig:dist_trace} shows this scheme as a sample distributed trace where the main shard, at the top, performs the majority of compute within each net.

Due to the massive size of embedding tables, a single table can also be partitioned to multiple shards, as shown in Figure~\ref{fig:sparsenn_shards}.
In such a case, the sparse feature IDs are split and sent to the appropriate RPC operator based on a hashing function.
This is implemented by partitioning embedding table rows with a simple modulus operator across shards.
The serving infrastructure imposes the constraint that graph cycles cannot exist between shards, so each shard is stateless to avoid further complexity.
This restriction also provides greater flexibility in a serving environment, where shards may fail and need to restart or replicas may be added.

\subsubsection{Serving Shards}
Distributed inference requires an additional special remote procedure call (RPC) operator, which replaces subnets in the main shard and invokes its respective \remoteshard{}, as shown in ~Figure~\ref{fig:sparsenn_shards}.
This scheme enables straightforward scale-out style support for large models.
An inference request gets sent to a server with the main shard loaded, and when an RPC op is encountered, a subsequent request to the appropriate \remoteshard{}s is issued.
Inference serving on all shards is comprised of an RPC service handler, such as \fbthrift ~or gRPC, and a machine learning framework, such as Caffe2, PyTorch, or TensorFlow~\cite{fbthrift, pytorch, tensorflow}.
Each shard runs a full service handler and ML framework instance.

This distributed architecture supports the same replication infrastructure used in non-distributed inference.
Shards are replicated based on their load and the resource needs of large-scale deployment.
For example, a shard that requires more compute resources to meet QPS requirements will have copies \textit{replicated} and deployed on its cluster via a cluster-level hypervisor.
The advantage to both main and \remoteshard{}s running a full service stack is that they can be replicated independently. 
Each individual request can then be processed by a different combination of machines than a previous request, further motivating the requirement of stateless shards.
The replication infrastructure is not enabled in our experiments because an isolated set of servers is used for in-depth characterization and analysis of per-request overheads.
A discussion of projected impacts and interactions of distributed inference on replication is included in Section~\ref{sec:data_center}.

\subsection{Capacity-Driven Model Sharding}
\label{sec:dist_inf_sharding}
Optimal model sharding is a challenging systems problem, due to the number of configurations and varying optimization objectives.
While this problem has been studied in the context of training, the inference context presents new challenges.
Training must contend with the accuracy implications of mini-batch sizing and parameter synchronization, and is targeted at maximizing throughput on a single instance that's fed a large dataset~\cite{demystifying_parallel}.
Data-center inference, in contrast, must contend with varying request rates and sizes under strict latency constraints, and the impact of model replication to meet those requirements.
Furthermore, in the context of recommendation, memory footprint is dominated by embedding tables which are computationally sparse.
Model sharding for deep learning recommendation inference is motivated by the desire to enable huge models, and thus we consider this new challenge \textit{capacity-driven} sharding.

Due to the number of shard configurations, an exhaustive search for an optimal sharding scheme is intractable.
Instead, heuristics are used, which depend on the model architecture and include measurements or estimates of model capacity, compute resources, and communication costs.
Naively, the heuristic should aim to \textit{minimize} the number of shards due to the additional compute and network resources consumed with more shards.
We study the resulting impact of this assumption in Section~\ref{sec:results}.
The heuristics are guided by the following observations:
\begin{enumerate}
	\item Embedding tables used in sparse layers contribute to the vast majority ($>97\%$) of model size.
	\item Sparse layers are memory and communication bound, while dense layers are compute bound.
	\item Existing server infrastructure cannot support memory requirements of sparse layers but can support the compute and latency requirements of dense layers.
	\item Deep recommendation model architectures, as in Figure~\ref{fig:dist_sparsenn}, can execute sparse operators in parallel, which feed successive dense layers.
\end{enumerate}

Because the dense layers do not benefit from additional compute or parallelism provided by distributed inference, our sharding strategies are constrained to only move sparse layers to remote shards. 
This is specific to the architecture of deep recommendation models.
Figure~\ref{fig:dist_sparsenn} shows that, after embedding table lookup and pooling, all of the resulting embedding vectors and dense features are combined in a series of feature interaction layers.
Placing the interaction layer on its own shards results in an undesirable increase of communication and lessening of compute density.
Placing this layer with existing \remoteshard{}s also increases communication and violates the stateless shard constraint.
Alternate architectures \emph{may benefit} from sharding interaction layers.
Consider a model architecture that has dedicated feature interaction layers for specific sets of embedding tables.
Such an architecture could shard those feature interaction layers with their respective sparse layers and indeed see performance benefits.
While an intriguing model architecture, we are limited to models that are currently available, thus this work chooses to focus on the more traditional model in Figure~\ref{fig:sparsenn}.

Thus, sharding the sparse layers (1) directly addresses capacity concerns of the largest parameters,
(2) effectively parallelizes sparse operators which otherwise execute sequentially, and
(3) enables better resource provisioning by isolating the communication- and compute-bound portions of inference.
Note that inference is not traditionally operator-parallel because operators do not typically produce enough work to offset scheduling overheads.
Extra computing cores are instead utilized by increasing batch-level parallelism.

Three \textit{sharding strategies}, using the above heuristic, are evaluated in this work (Table~\ref{table:sharding_summary}).
These strategies address the new and distinct challenges of huge deep learning recommendation models: the varied number and size of embedding tables, non-uniformity of sparsity, and the challenge of the real-time serving environment.
Two trivial cases of (1) non-distributed inference, or singular, and (2) a single shard with all embedding tables are also presented in Table~\ref{table:sharding_summary}.

\begin{table}[t]
	\caption{Sharding Strategy Summary}
	\begin{center}
		\begin{tabularx}{\linewidth}{ |X|X| }
			\hline
			\multicolumn{1}{|c|}{\textbf{Sharding Strategy}} & \multicolumn{1}{c|}{\textbf{Notes}} \\
			\hhline{|==|}
			Singular & Distributed inference disabled. \newline 
			Entire model loaded on one server. \\
			\hline
			1 Shard - All Tables & Only one \remoteshard{} with all embedding tables. \\
			\hline
			\{2, 4, 8\} Shards Capacity-balanced & Table placement ensures similar total embedding table size per shard \\
			\hline
			\{2, 4, 8\} Shards Load-balanced & Table placement ensures similar pooling work per shard  \\
			\hline
			\{2, 4, 8\} Shards Net-specific bin-packing (NSBP) & Tables are grouped by ML net, and packed into shards until a size limit is reached. Tables larger than this limit are effectively given an entire shard.   \\
			\hline
		\end{tabularx}
		\label{table:sharding_summary}
	\end{center}
\end{table}

\subsubsection{Capacity-balanced}
An intuitive strategy is to spread embedding tables evenly across many shards.
\textit{Capacity-balanced} sharding ensures that each \remoteshard{} has the same memory requirements.
This serves to minimize the number of shards, with the goal to achieve the least compute \emph{resource} overhead for a singly served model.

\subsubsection{Load-balanced}
Each sparse feature is represented as a multi-hot encoded vector, which transforms to a multi-index lookup into the embedding table and a final pooling operation.
Because the expected number of lookups per table is dependent on specific feature and how often it appears in request inputs, capacity-balanced sharding may result in imbalanced shards that perform signficantly different amounts of work.
Additionally, the number of lookups is proportional to the network bandwidth used to send table indices.
This aggregate imbalance can cause certain shards to become a critical path bottleneck, degrading latency compared to the load-balanced sharding configuration. 

The distributed trace shown in Figure~\ref{fig:dist_trace} demonstrates how one \remoteshard{} may become on the critical path, albeit unpredictable variance in network latency must also be considered.
Remote shard 1 and 2 are queried asynchronously, in parallel.
Because remote shard 1 performs significantly more work, more latency overhead is incurred. 
To reduce the likelihood of one shard consistently increasing latency, the \textit{load-balancing} strategy places embedding tables based on their \textit{pooling factor}, or expected number of lookups.
The pooling factor is estimated by sampling 1000 requests from the evaluation dataset and observing the number of lookups per table.

\subsubsection{Net-specific bin-packing}
Recommendation models frequently separate the user features and content/product features into distinct nets to more effectively batch user-content pairs together~\cite{fb_rec_hpca}.
In the models chosen for this work, output of the user net is fed into the content/product net, so they must be executed sequentially.
An example of this is shown in Figure~\ref{fig:dist_trace}, where Net 2 is dependent on Net 1.
Separate RPC operators will be called when each net is executed to access their respective embedding tables.
If tables are not grouped by net, \emph{the same shard may be accessed multiple times per batch, for each net}.
This is undesirable because 
1) in Figure~\ref{fig:dist_trace}, three RPC operators are invoked, compared to two (one RPC per shard),
and 2) as servers are replicated in a data center environment to handle increased requests, tables for both nets will be duplicated regardless of which net receives more inputs.

This is particularly undesirable if, in Figure~\ref{fig:dist_trace}, server replication of remote shard 1 is triggered by high compute characteristic of Net 1's pooling. All of Net 2's embedding tables, which may be 10s of GBs sharded, will consume additional, under-utilized memory resources.
To account for this and target more efficient resource usage, a \textit{net-specific bin-packing} (NSBP) strategy is evaluated, which first groups tables by net, and then packs them into bins based on a given size constraint.
To reduce data-center resources incurred \textit{during} sharding, each bin starts out as the existing sparse parameter servers used during training. 
If a parameter server's \textit{bin} is already full, it is considered a full shard.
This reduces network bandwidth and compute orchestration required for sharding.

\begin{figure}[t]
	\centering
	\includegraphics[width=\linewidth]{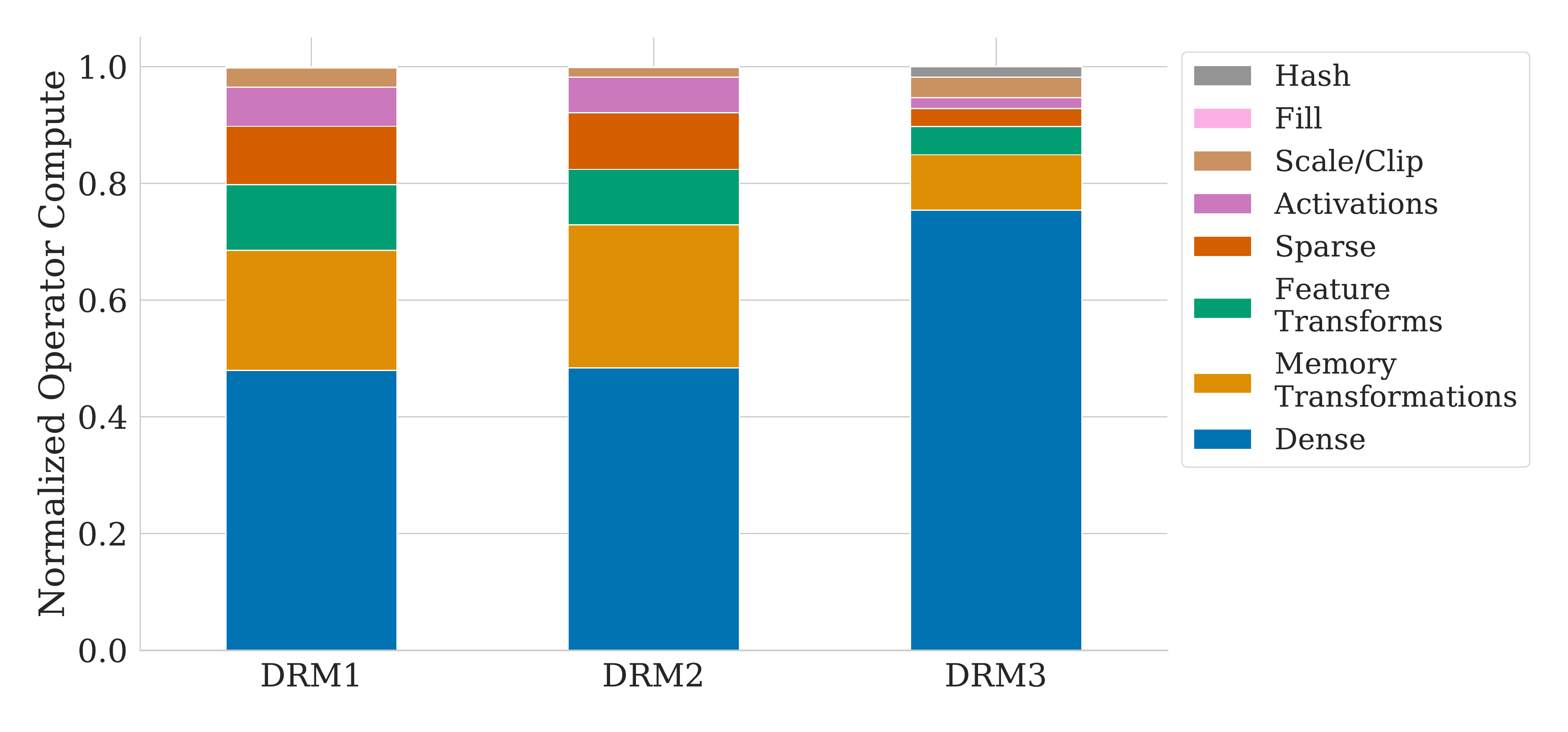}
	\caption{\changed{Operator compute attributions for \RMone{}, \RMtwo{}, and \RMthree{}. The models are architecturally similar, but have different characteristics. \RMone{} and \RMtwo{} differ sparse features and average request size. \RMthree{} is more sparse than \RMone{} and \RMtwo{}}}
	\label{fig:op_stack}
\end{figure}

\subsection{Distributed Inference Implementation}
\label{sec:dist_inf_implementation}
We present an in-depth description of the customized open-source frameworks used in this work.
This provides the reader with a concrete implementation to frame our results and provides guidance for implementation with other frameworks. 
For this work, distributed inference is built on top of a highly customized variant of \fbthrift ~and \textit{Caffe2}, however the methodology is generalizable to any RPC or ML framework~\cite{fbthrift, pytorch}.
While PyTorch has absorbed and succeeded Caffe2 as the state-of-the-art moving forward, the Caffe2 infrastructure used in this work is shared between both frameworks. 
\fbthrift ~serves ranking requests by loading models and appropriately splitting received requests to inference batches to the appropriate net.
A modified version of Caffe2 is used that includes RPC operators that issues \fbthrift ~requests.
The intermediate requests to \remoteshard{}s are routed via a universal service discovery protocol.
All inter-server communication occurs through standard TCP/IP stack over Ethernet.
The model is transformed for distributed inference after training.
A custom partitioning tool employs a user-supplied configuration to group embedding tables and their operators, insert RPC operators, generate new Caffe2 nets, and then serialize the model to storage.

\begin{figure}[t]
	\centering
	\includegraphics[width=\linewidth]{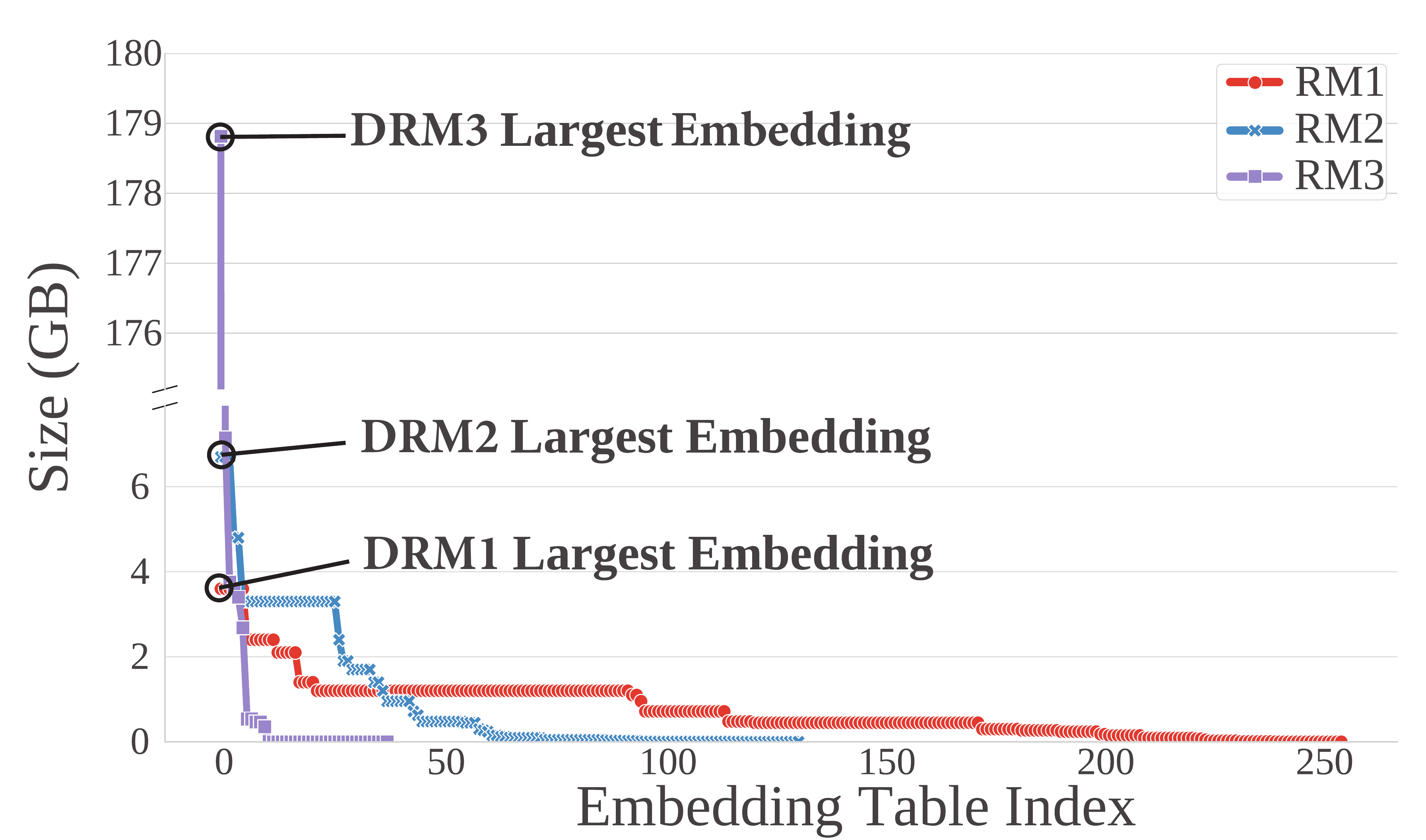}
	\caption{Embedding Table Size Distribution. \RMthree{}'s size is dominated by a single large table, compared to the heavier tails exhibited by \RMone{} and \RMtwo{}.}
	\label{fig:emb_tbl_size}
\end{figure}

\section{Cross-Layer ML Operator- and Communication-Aware Characterization}
\label{sec:tracing_framework}
The performance of distributed inference is determined by choices at multiple layers of the system, in particular, the data-center service level (scheduling, discovery, and networking), the machine learning framework-level, and then the machine learning operators themselves. Measuring a workload across these layers is important for understanding overheads, attributing costs, targeting components for optimization, and making high-level system design decisions.  
Because no profiling tools exist to perform such a cross-layer characterization, we built a custom cross-layer distributed tracing framework for measuring distributed inference workloads.

\begin{table*}[t]
	
	\centering
	\caption{Sharding Results for \RMone{}. Each column is a different sharding configuration. Each row is a static attribute. The bracketed elements $[N]: value$ represent the $N^{th}$ shard's attribute value for that configuration.}
	\ssmall
	\begin{center}
		\begin{tabularx}{\textwidth}{ | X | X | X | X | X | X | X | X | X | X | X |  }
			\hline
			& \multirow{2}{*}{\textbf{1-shard}} & \multicolumn{3}{c|}{\textbf{Load-balanced}} & \multicolumn{3}{c|}{\textbf{Capacity-balanced}} &  \multicolumn{3}{c|}{\textbf{Net-Specific Bin-Packed}} \\
			\cline{3-11}
			&& 2 & 4 & 8 & 2 & 4 & 8 & 2 & 4 & 8 \\
			\hhline{|===========|}
			Capacity (GiB) & [1]: 194.05 & [1]: 89.38 \newline [2]: 104.67 & 
			[1]: 40.94 \newline [2]: 60.76 \newline [3]: 44.16 \newline [4]: 48.18 & 
			[1]: 28.87 \newline [2]: 29.82 \newline [3]: 18.23 \newline [4]: 21.0 \newline [5]: 20.5 \newline [6]: 26.35 \newline [7]: 23.44 \newline [8]: 25.85 & 
			[1]: 97.03 \newline [2]: 97.03 &
			[1]: 48.52 \newline [2]: 48.51 \newline [3]: 48.51 \newline [4]: 48.51 &
			[1]: 24.25 \newline [2]: 24.25 \newline [3]: 24.25 \newline [4]: 24.25 \newline [5]: 24.25 \newline [6]: 24.25 \newline [7]: 24.25 \newline [8]: 24.25 &
			[1]: 33.58 \newline [2]: 160 &
			[1]: 55.89 \newline [2]: 48.22 \newline [3]: 55.89 \newline [4]: 33.58 &
			[1]: 27.93 \newline [2]: 5.649 \newline [3]: 27.95 \newline [4]: 27.94 \newline [5]: 27.94 \newline [6]: 27.95 \newline [7]: 27.95 \newline [8]: 20.28 \\
			\hline
			Embedding Tables &
			[1]: 257 &
			[1]: 125 \newline [2]: 132 &
			[1]: 63 \newline [2]: 67 \newline [3]: 63 \newline [4]: 64 &
			[1]: 33 \newline [2]: 31 \newline [3]: 33 \newline [4]: 32 \newline [5]: 32 \newline [6]: 31 \newline [7]: 32 \newline [8]: 33 &
			
			[1]: 121 \newline [2]: 136 &
			[1]: 62 \newline [2]: 74 \newline [3]: 60 \newline [4]: 61 &
			[1]: 43 \newline [2]: 31 \newline [3]: 30 \newline [4]: 31 \newline [5]: 31 \newline [6]: 31 \newline [7]: 31 \newline [8]: 29 &
			
			[1]: 72 \newline [2]: 185 &
			[1]: 28 \newline [2]: 106 \newline [3]: 51 \newline [4]: 72 &
			[1]: 42 \newline [2]: 30 \newline [3]: 18 \newline [4]: 11 \newline [5]: 43 \newline [6]: 27 \newline [7]: 27 \newline [8]: 59
			\\
			\hline
			Estimated Pooling Factor & 
			[1]: 138943.1 & [1]: 69471.6 \newline [2]: 69471.5 &
			[1]: 34735.9 \newline [2]: 34735.7 \newline [3]: 34735.7 \newline [4]: 34735.8 &
			[1]: 17367.4 \newline [2]: 17368.3 \newline [3]: 17367.5 \newline [4]: 17368.2 \newline [5]: 17368.1 \newline [6]: 17367.6 \newline [7]: 17368.4 \newline [8]: 17367.6 & 
			
			[1]: 65852.3 \newline [2]: 73090.8 &
			[1]: 28392.9 \newline [2]: 35719.4 \newline [3]: 38163.4 \newline [4]: 36667.4 &
			[1]: 24656.9 \newline [2]: 12067.0 \newline [3]: 29746.4 \newline [4]: 6310.7 \newline [5]: 11825.0 \newline [6]: 14050.1 \newline [7]: 21075.8 \newline [8]: 19211.2 &
			
			[1]: 126652.7 \newline [2]: 8010.7 &
			[1]: 1859.9 \newline [2]: 3540.5 \newline [3]: 2610.3 \newline [4]: 126652.7 &
			[1]: 77103.4 \newline [2]: 49549.3 \newline [3]: 1034.2 \newline [4]: 781.3 \newline [5]: 2017.9 \newline [6]: 1372.3 \newline [7]: 1532.2 \newline [8]: 1272.8 \\
			\hline
		\end{tabularx}
		\label{table:shard_results}
	\end{center}
\end{table*}

\subsection{Workload Capture via Distributed Tracing}
Distributed tracing is a proven technique for debugging and performance investigations across inter-dependent services\cite{dapper, canopy}.
We leverage this technique to investigate the performance impact of the distributed inference infrastructure.
Custom instrumentation was added at multiple layers in the production service stack to provide a complete view of costs for request/response serialization, RPC service boilerplate setup, and model execution for each request.

\fbthrift ~and Caffe2 both provide instrumentation hooks to capture salient points in execution, leaving the service handler to be more intrusively modified, albeit still lightweight.
Source instrumentation was chosen over other binary instrumentation approaches because it's lighter weight and can interface with \fbthrift's \emph{RequestContext} abstraction to propagate contextual data for distributed tracing.
At each trace point, metadata specific to the layer and a wall-clock timestamp are logged to a lock-free buffer and then asynchronously flushed to disk.
Wall-clock time is desirable because its ordering helps achieve a useful trace visualization.
Furthermore, most spans are small and sequential, enabling wall-clock time as a proxy for CPU time.
Per-shard CPU time for each request is also logged to verify this claim.
The trace points are then collected and post-processed offline for overhead analysis and to reconstruct a visualization of events, resembling Figure~\ref{fig:dist_trace}. 

An example distributed trace is shown in Figure~\ref{fig:dist_trace}. 
Shards are separated by horizontal slices, with the main shard servicing requests located at the top.
The example trace represents a single batch, however it's typical for multiple batches to be executed in parallel depending on the number of ranking requests and the configurable batch size.
Furthermore, the ordering of operators is typical for the recommendation models studied in this work.
Operators are scheduled to execute sequentially--unless specifically asynchronous like the RPC ops--because other cores are utilized via request- and batch-level parallelism.
The initial dense layers are first executed, followed by the sparse lookups, and then the later feature interaction and top dense layers.

As discussed in Section\ref{sec:dist_inf_serving}, each \remoteshard{} is a full RPC service for deployment flexibility,
and as such incurs service-related overheads compared to local, in-line computation.
Figure~\ref{fig:dist_trace} demonstrates  that latency overhead in each \remoteshard{} is any time \textit{not} spent executing SLS operators.
In the non-distributed case, the SLS ops are directly executed in the main shard.
The extra latency is attributed to the network link, extra (de)serialization of inputs and outputs, and time spent preparing and scheduling the Caffe2 net. 
Despite such overheads, we also see that the asynchronous nature of the distributed computation enables more parallelization of the sparse operators.
\textit{An important takeaway from this characterization is the trade-off between increased parallelization to reduce latency overhead and decreased parallelization to reduce compute and data-center resource overhead}. 
Furthermore, this trade-off is specific to each model.

\subsection{Cross-Layer Attribution}
Compared to simple, operator-based counters for compute or end-to-end (E2E) latency, capturing a cross-layer trace provides a holistic view of compute and latency.
The example flow in Figure~\ref{fig:dist_trace} shows compute overheads from RPC serialization and deserialization and request/response processing of the software infrastructure.
Additional compute is incurred in each thread's scheduling and book-keeping of asynchronous RPC operators.
Simple end-to-end latency overhead is easily measured at the main shard.
However, because each \remoteshard{} is executed in parallel, attribution of latency overheads is more complex and involves overlap between \remoteshard{} requests.
To simplify analysis, the slowest asynchronous \remoteshard{} request, per main shard request, is used for latency breakdown.
The network-, serialization-, and RPC service- latencies in the associated \remoteshard{} are used.
Because the clocks on disparate servers will be skewed, network latency is measured as the difference between the outstanding request measured at the main shard and the end-to-end service latency measured at the \remoteshard.

\section{Experimental Methodology and Workloads}
\label{sec:methodology}
We evaluated production-grade distributed serving software infrastructure to quantify the impacts on compute requirements, latency, and effectiveness of sharding.
The purpose of this study was to evaluate the practicality of distributed inference as a means to support huge deep recommendation models and provide a basis for further systems exploration with this novel workload.
In this section, we provide descriptions of the data-center deep learning recommendation models chosen for this study and the underlying hardware platform.

\subsection{Recommendation Models}
\label{sec:model_test_description}
Three DLRM-like models, referred to as \RMone{}, \RMtwo{} and \RMthree{}, were used that span a range of large model attributes, such as varying \textit{input features} and \textit{embedding table characteristics}.
The D-prefix is to contrast the \emph{distributed} models to the specific models discussed in recent deep learning recommendation works~\cite{guptadeeprecsys_isca20, fb_rec_hpca, fb_dlrm}.
The DRM* models are a subset of many possible, evolving models and chosen as early candidates for distributed inference given their sparse feature characteristics.
The goal of studying these models is to present a basis for evaluating overheads of distributed inference via our cross-layer distributed trace analysis, and they should not be interpreted as canonical benchmark models such as those included in the MLPerf benchmark suite~\cite{mlperf-inference}.

Figure~\ref{fig:op_stack}, Figure~\ref{fig:emb_tbl_size}, and Table~\ref{table:shard_results} demonstrate the variety within large neural recommendation models.
\RMone{} and \RMtwo{} have two nets, each having their own respective sparse features, while \RMthree{} has a single net.
Because real, sampled requests were used for each model, the inference request size and corresponding compute and latency also varied between models.
All parameters were uncompressed as single-precision floating point.
Section~\ref{sec:compression} discusses the impact of compression.
Per-operator group attributions for each model are shown in Figure~\ref{fig:op_stack}, as a simple mean average across all sampled requests for the non-distributed model.
\RMone{} and \RMtwo{} are the most similar architectures, reflected in Figure~\ref{fig:op_stack}.
\changed{Compared to \RMthree{}, \RMone{} and \RMtwo{} have a more complex structure evidenced by additional tensor transform costs.
More relevant to this work, sparse operators consume a much larger portion of all operators compared to \RMthree{}.}
Specifically, sparse operators contribute to 9.7\%, 9.6\%, and \changed{3.1\%} of all operator time, in \RMone{}, \RMtwo{}, and \RMthree{} respectively.
Despite their low proportional compute, the sparse operators account for  \textgreater97\% of model capacity in \RMone{} and \RMtwo{}, and \changed{\textgreater99.9\% of capacity for \RMthree{}}.
Embedding tables larger than a given threshold were scaled down by a proportional factor to fit the entire model on a single 256GB server.
This provided a straight comparison of compute and latency overheads across all sharding strategies listed in Table I, including singular. The original data-center scale models are many times larger.

The distribution of embedding tables sizes within each model is shown in Figure~\ref{fig:emb_tbl_size}.
\RMone{} was sized to 200GB with 257 embedding tables and the largest table is at 3.6GB.
\RMtwo{} was proportionately sized to 138GB with 133 embedding tables and the largest table at 6.7GB.
\changed{Lastly, \RMthree{} was sized to 200GB with 39 embedding tables and the largest table at 178.8GB.}
\RMone{} and \RMtwo{} demonstrate a long tail of embedding table size compared to \RMthree{}, explaining the additional sparse operator cost shown in Figure~\ref{fig:op_stack}.
Comparatively, \RMthree{} is dominated by a single large table.
The embedding tables are representative versions of the tables discussed in prior deep learning recommendation work~\cite{fb_dlrm, fb_rec_hpca, mlperf-inference}.

For \RMone{} and \RMtwo{}, ten sharding configurations were evaluated.
Sharding strategy results for \RMone{} are described in Table~\ref{table:shard_results}.
For the load-balanced configuration, per-shard capacities varied up to 50\% compared to capacity-balanced where each shard is the same total capacity.
For the capacity-balanced configuration, per-shard estimated load varied up to 371\%, between shards 4 and 3 with eight shards, compared to load-balanced where each shard had the same total pooling factor.
Lastly, the net-specific bin-packing (NSBP) strategy restricted each shard's tables to a single net.
This is most noticeable at the 2-shard configuration, where each net is placed on its own shard.
Shard 2 consumes \textbf{4.75}$\times$ as much memory capacity as Shard 1, yet is estimated to perform just \textbf{6.3}\% of Shard 1's compute work.

\RMthree{} is only sharded with NSBP, and not capacity-balanced or load-balanced strategy, due to existing technical challenges of sharding huge tables.
Because it is dominated by a single large table, for each additional shard the largest table is further split, while the smaller tables remained grouped as one shard.
The dominating table has a pooling factor of \textbf{1}, thus only one of the shards spanning the table will be accessed.
For example, given four \remoteshard{}s, the largest table is partitioned into three shards and the remaining tables grouped together into one shard.
Each inference, only two shard would be accessed: one for the sharded large table and one for the smaller tables.

\begin{figure*}[t]
	\centering
	\begin{subfigure}{\linewidth}
		\includegraphics[width=\linewidth]{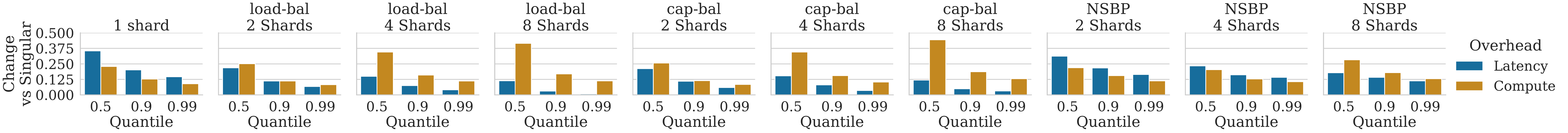}
		\caption{\RMone{}}
	\end{subfigure}

	\begin{subfigure}{\linewidth}
		\includegraphics[width=\linewidth]{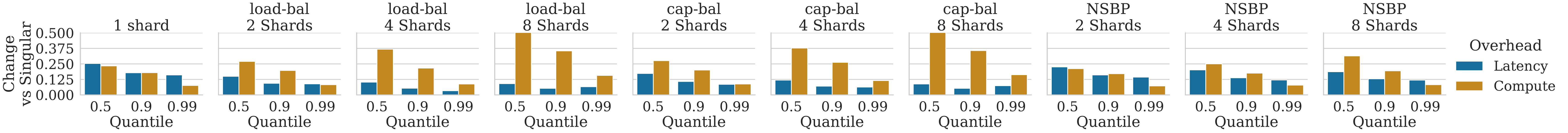}
		\caption{\RMtwo{}}
	\end{subfigure}

	\caption{P50, P90, and P99 latency and compute overheads compared to the singular model. Latency and compute have an
		inverse relationship due to additional asynchronous scheduling overheads. Increasing shards effectively increases parallelization.}
	\label{fig:overheads}
\end{figure*}

\begin{figure}[t]
	\centering
	\includegraphics[width=\linewidth]{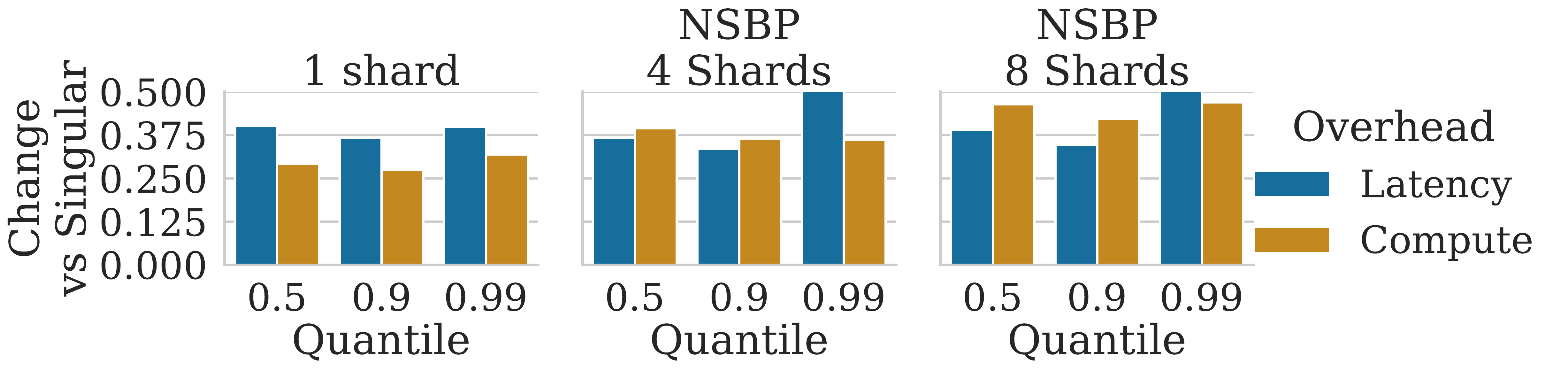}
	\caption{\RMthree{} P50, P90, and P99 latency and compute time overheads compared to singular. Increasing shards does not increase parallelization as compared to \RMone{} and \RMtwo{}.}
	\label{fig:overheads_rm3}
\end{figure}

\begin{figure*}[t]
	\centering
	\begin{subfigure}{.49\textwidth}
		\centering
		\includegraphics[width=\linewidth]{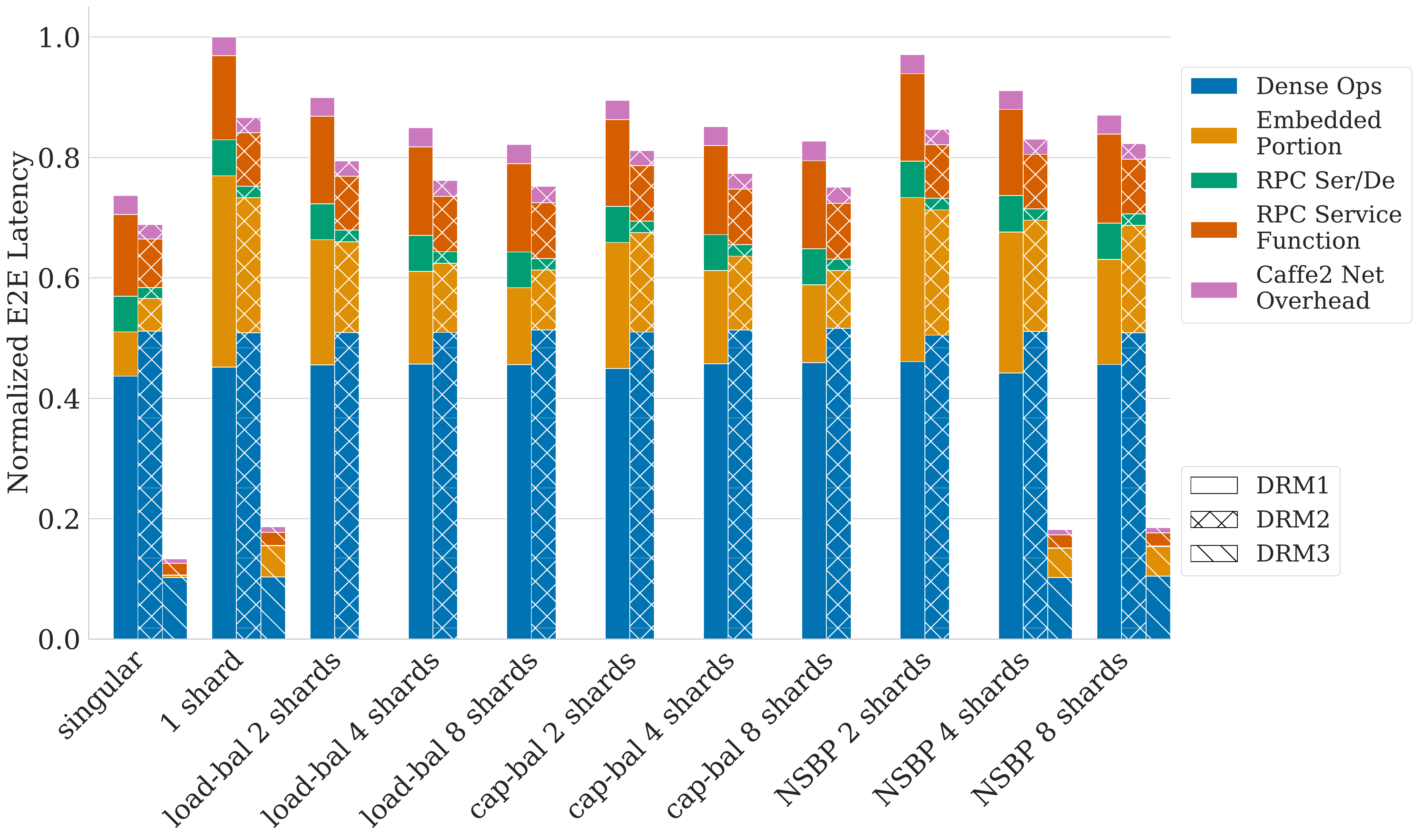}
		\caption{Total E2E Latency Stack, measured at the main shard.}
		\label{fig:latency_attribution_e2e}
	\end{subfigure}%
	\begin{subfigure}{.49\textwidth}
		\centering
		\includegraphics[width=\linewidth]{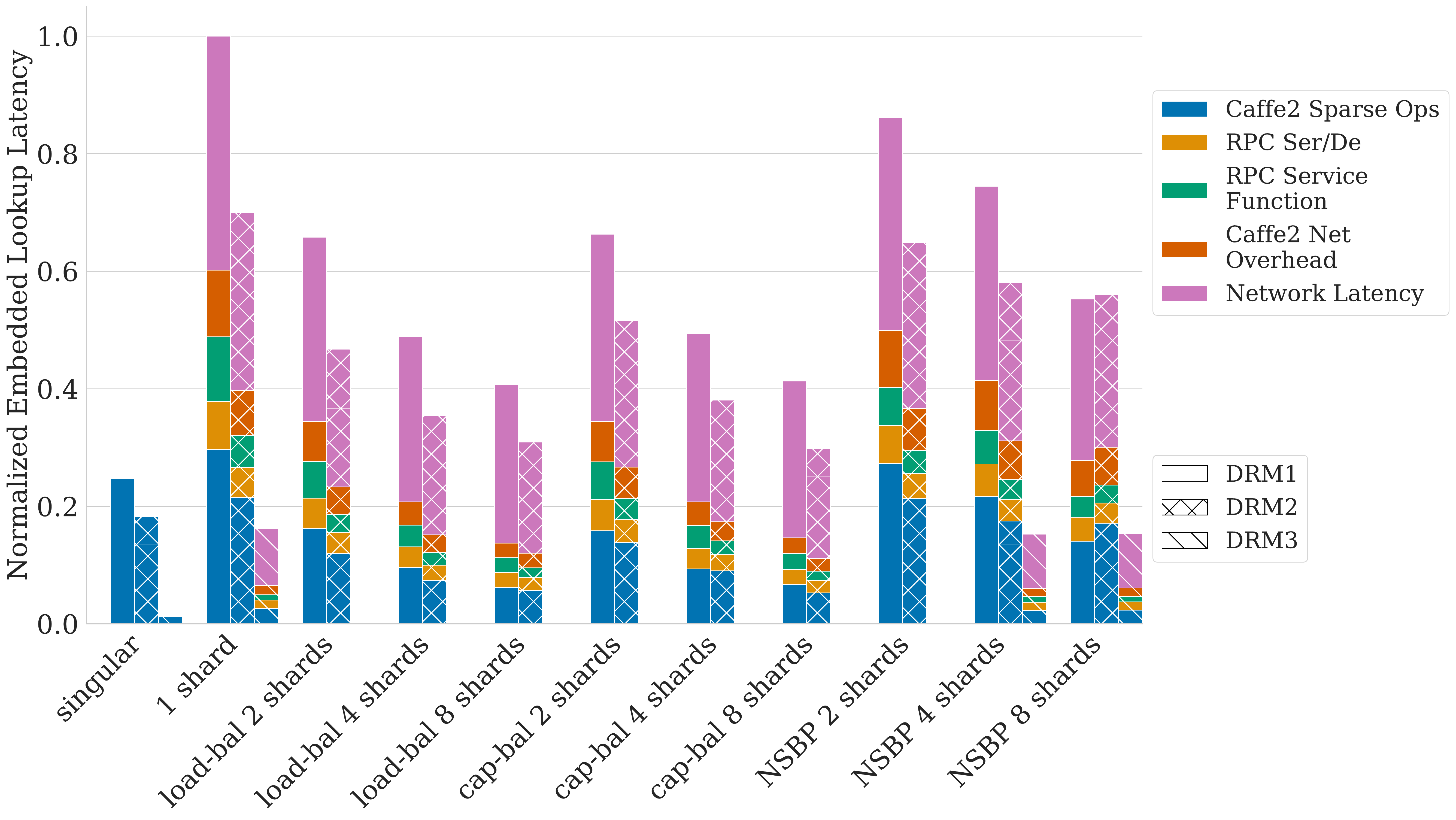}
		\caption{Embedded Portion Latency Stack, for the bounding shard.}
		\label{fig:latency_attribution_sls}
	\end{subfigure}%
	\caption{P50 Latency Attribution by Sharding Strategy. Latencies are normalized to the \emph{highest} latency configuration to contrast models. Increasing parallelism with more shards reduces aggregate latency
		overheads, but constant network latency eventually dominates. Not all strategies parallelize to the same degree.} 
	\label{fig:latency_attribution}
\end{figure*}

\begin{figure}[t]
	\centering
	\includegraphics[width=\linewidth]{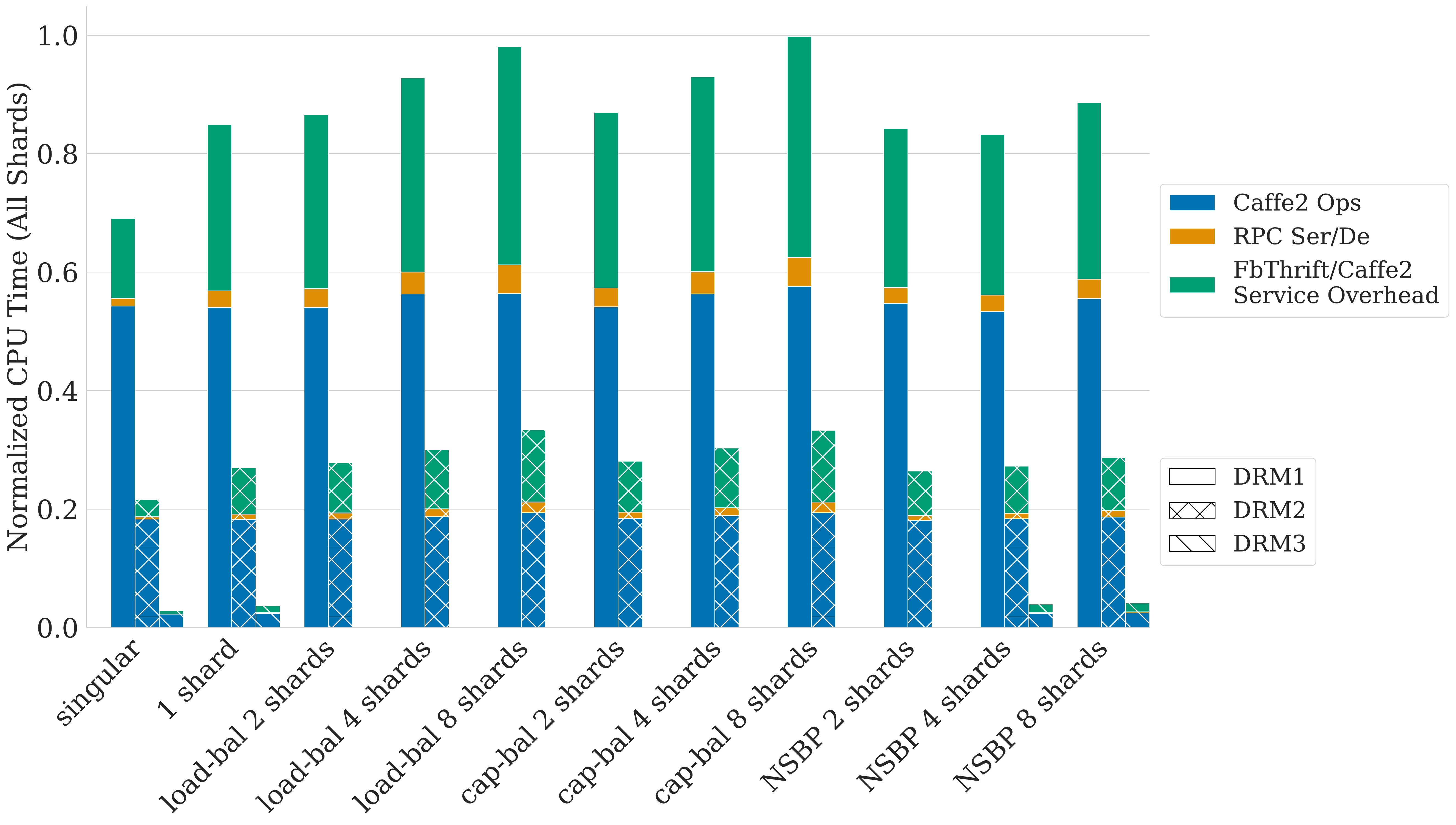}
	\caption{P50 Aggregate CPU Time Stack by sharding config. CPU Time is normalized to highest configuration.}
	\label{fig:compute_attribution_breakdown}
\end{figure}

\subsection{Test Platform}
Two classes of servers, representative of the data-center environment, were used for characterization.
\textit{\SCLarge{}} is representative of a typical large server in a data-center and has 256GB of DRAM and two 20-core Intel CPUs.
\textit{\SCSmall{}} is representative of a typical, more efficient web server and has 64GB of DRAM and two, slower clocked, 18-core Intel CPUs, and less network bandwidth than \SCLarge{}. 
Because of the limited DRAM available on \SCSmall{}, only a subset of configurations could be tested on this platform.
The majority of results discussed in Section~\ref{sec:results} were collected on \SCLarge{} platforms for an apples-to-apples comparison with the non-distributed models.
A discussion on the impact of server platform follows in Section~\ref{sec:data_center}.

All inference experiments were run on CPU platforms, as discussed in Section~\ref{sec:background}.
The reserved, bare-metal servers were located in the same data centers as production recommendation ranking and utilized the same intranet.
Their locations within the data center was representative of a typical inference serving tier.
Shards were assigned to unique servers and not co-located.
Shard-to-server mappings were randomized across repeated trials.
Recommendation ranking requests were sampled from production servers and replayed on the test infrastructure;
a database of de-identified requests were sampled evenly across a five-day time period in order to capture any diurnal behavior within the requests.
The production replayer then pre-processed and cached the requests before sending them to the inference servers using the same networking infrastructure used in online production ranking.

For the majority of experimental runs, batch sizes for inference were set to production defaults, where each batch represents a number of recommendation items to rank and is executed in parallel.
In Section~\ref{sec:results} requests were sent serially, to isolate inherent overheads.
In Section~\ref{sec:data_center_qps} requests were sent asynchronously, to simulate a higher QPS rate more representative of the production environment.
\section{Distributed Inference Sharding Analysis}
\label{sec:results}

In this section, we discuss the findings of the cross-layer distributed trace analysis.
We find that the design space is more nuanced than naively minimizing the number of shards, given hard latency targets and compute scale of the data-center environment.
Our primary takeaways are:
\begin{itemize}
	\item Increasing the number of shards can manage latency overheads incurred by the RPC operators by effectively increasing model parallelism.
	\item However, increased sharding also incurs substantial compute costs for each additional RPC, from service boilerplate and scheduling.
	\item Blocking requests sent serially, one-at-a-time, always perform worse in distributed inference across P50, P90, and P99 latencies, due to a simple Amdahl's law bound. Embedding lookups for \dlrm{}s do not comprise a large enough fraction of E2E latency to benefit from the increased parallelism.
	\item The load-balanced strategy did not significantly affect E2E latency compared the capacity-balanced strategy. The net-specific bin-packing (NSBP) sharding performed the worst, latency-wise, but the best compute-wise.
	\item Some models have insufficient work to parallelize and don't receive overhead reduction from additional shards.  
\end{itemize} 

The findings show distributed inference is a practical solution to serve the large, deep learning recommendation models, which the existing serving paradigm cannot support.
The findings also expose the need for targeted analyses of the network link between shards and the supporting software infrastructure for managing communication, given the little work performed on \remoteshard{}s.
The results suggest that an automatic sharding methodology is feasible, but requires sufficient profiling data given the variety in embedding table behavior and complex trade-offs of sharding strategies.

\begin{figure*}[t]
	\begin{subfigure}{.49\textwidth}
		\centering
		\includegraphics[width=\linewidth]{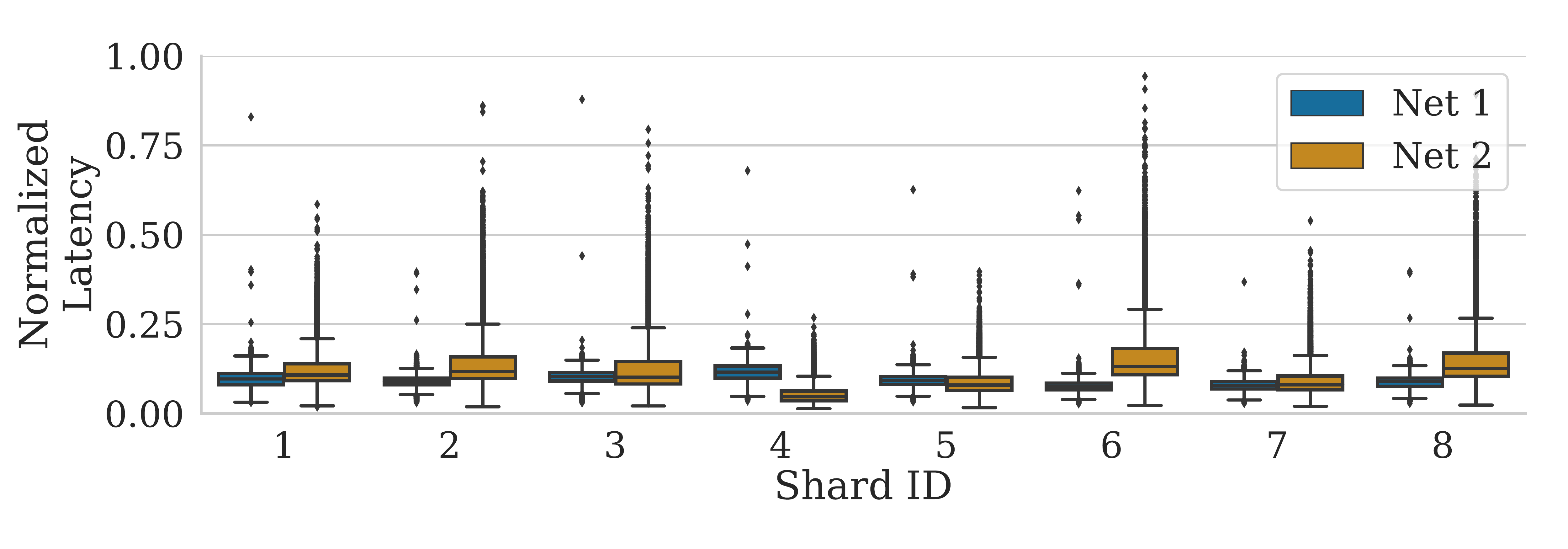}
		\caption{Load-balanced}
		\label{fig:latency_variability_remote_shards_rm1_load_bal}
	\end{subfigure}
	\begin{subfigure}{.49\textwidth}
		\centering
		\includegraphics[width=\linewidth]{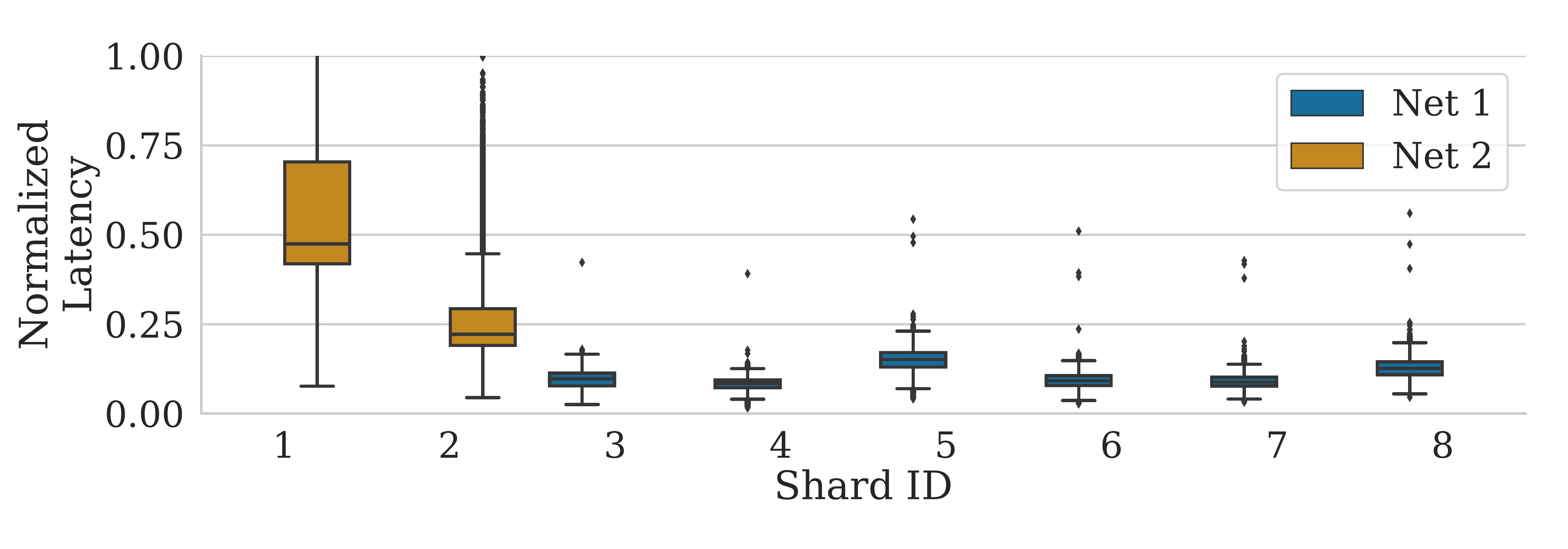}
		\caption{\changed{Net-specific Bin-packing}}
		\label{fig:latency_variability_remote_shards_rm1_nsbp}
	\end{subfigure}%
	\caption{\RMone{} per-shard operator latencies, by net w/ 8 \remoteshard{}s. Only co-locating tables within the same net has a large effect on latency.}
	\label{fig:latency_variability_remote_shards_pernet}
\end{figure*}

\begin{figure*}[t]
	\begin{subfigure}{.49\textwidth}
		\centering
		\includegraphics[width=\linewidth]{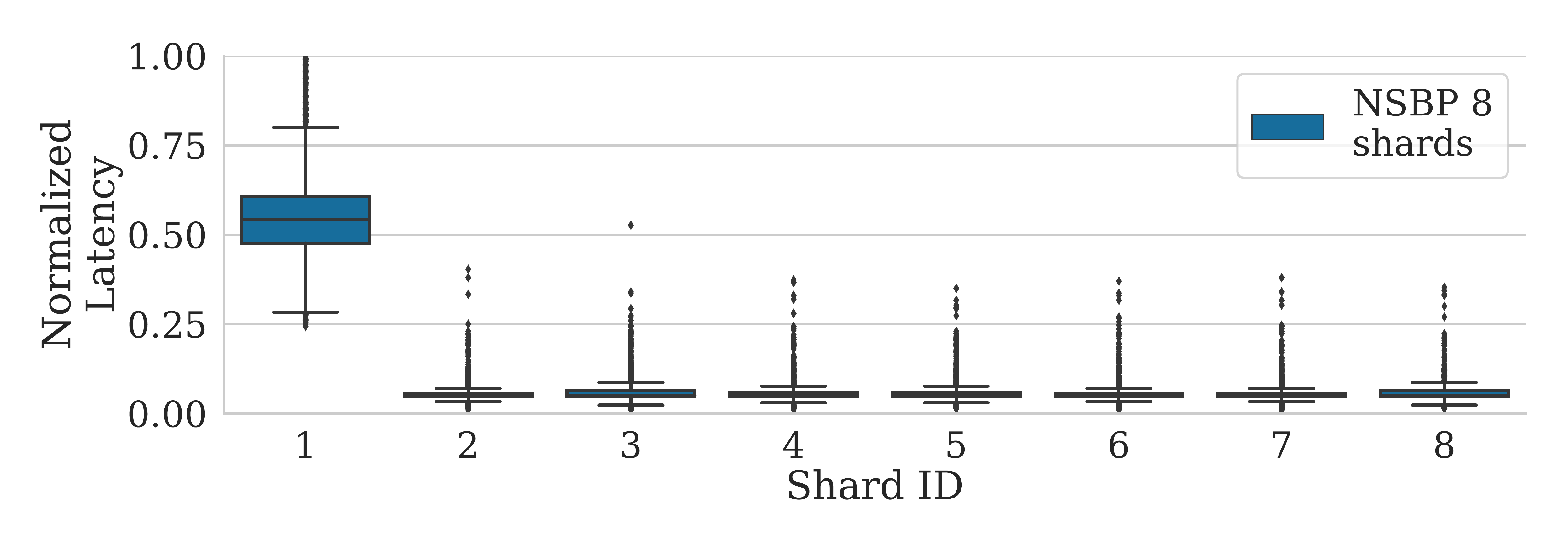}
		\caption{\RMthree{} Net-specific Bin-packing (NSBP). Shard 1 contains all
			tables except the largest, which is split across Shards 2-8. Each
			\RMthree{} inference makes one access to one of Shards 2-8.}
		\label{fig:rm3_per_shard}
	\end{subfigure}
	\qquad
	\begin{subfigure}{.49\textwidth}
		\centering
		\includegraphics[width=\linewidth]{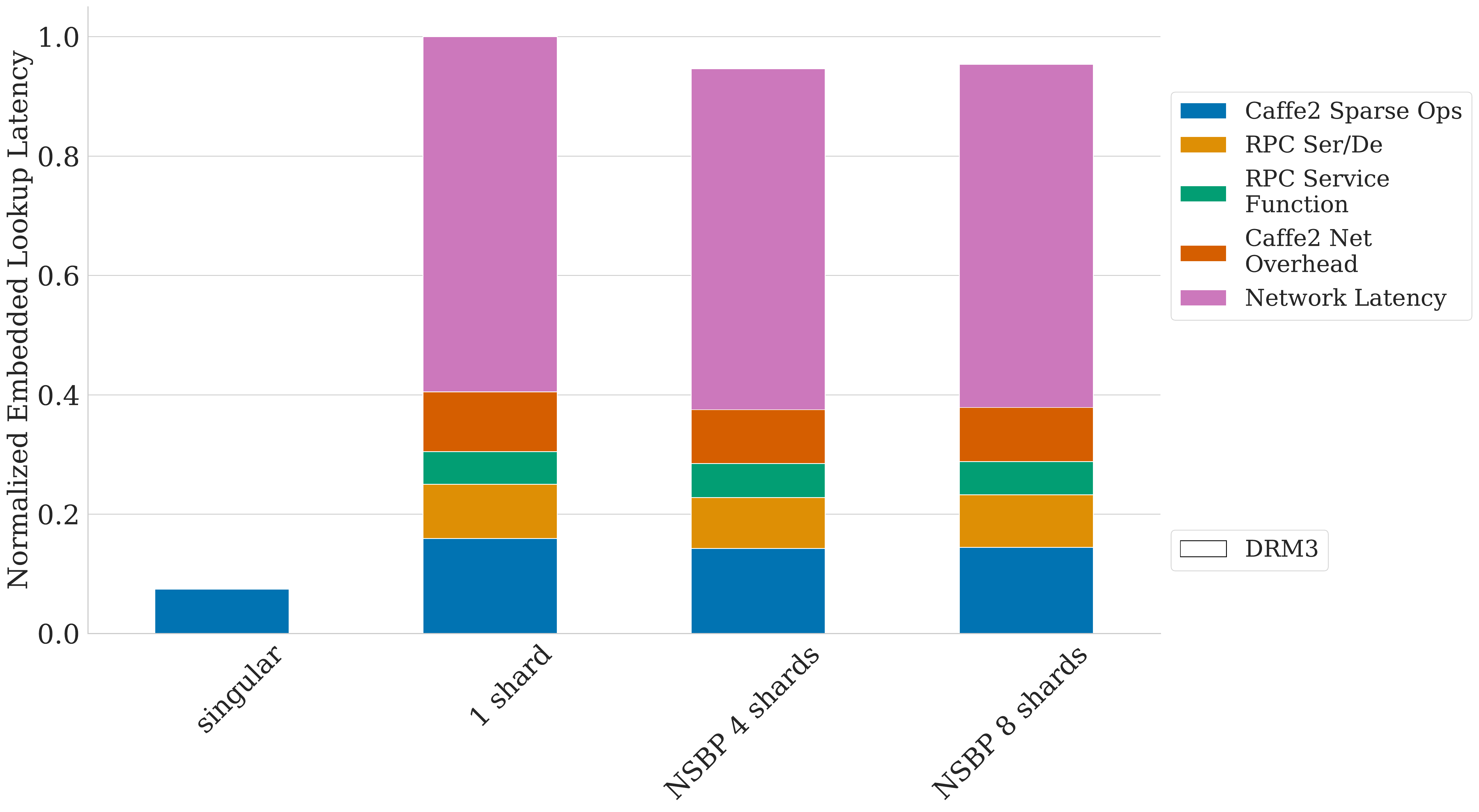}
		\caption{Embedded Portion Latency Stack.}
		\label{fig:rm3_latency_emb}
	\end{subfigure}%
	\caption{\changed{\RMthree{} per-shard operator latencies and embedded portion breakdown. \RMthree{}'s capacity is dominated by a single large embedding table. Increasing shards has no practical effect on latency.}} 
	\label{fig:rm3_specific}
\end{figure*}

\begin{figure}[t]
	\centering
	\includegraphics[width=\linewidth]{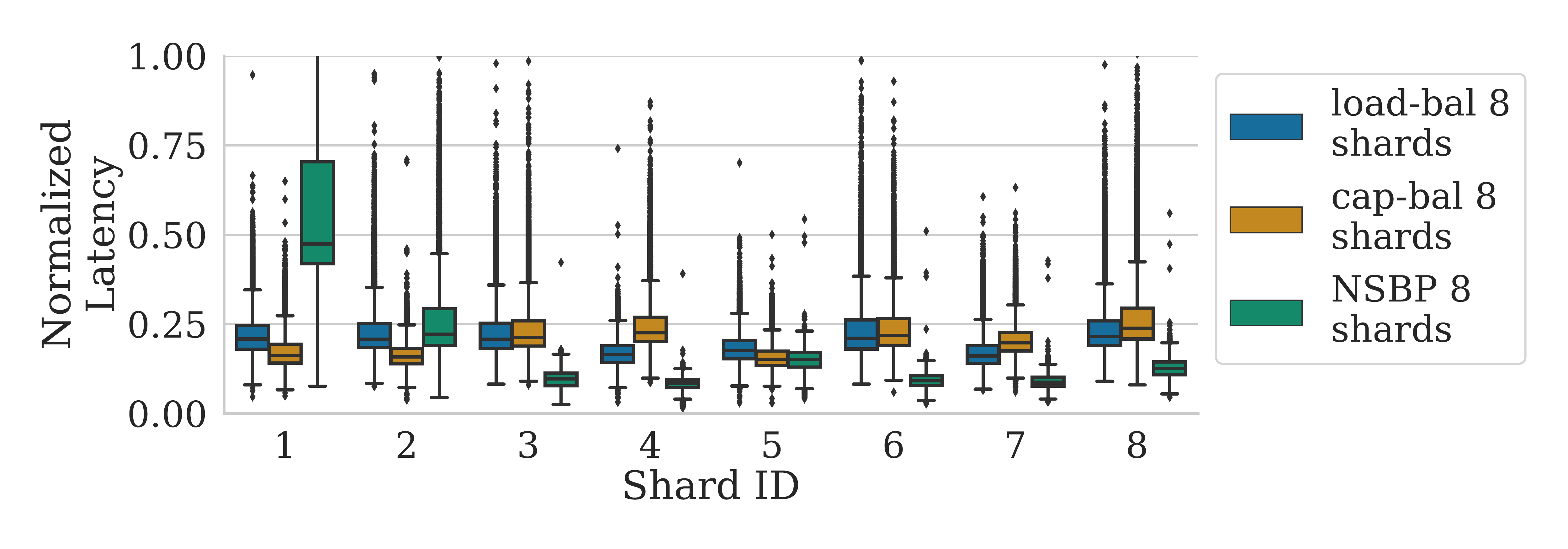}
	\caption{\RMone{} per-shard operator latencies, by sharding strategy w/ 8 \remoteshard{}s. Load-balanced does not substantially affect latency compared to capacity-balanced.}
	\label{fig:latency_variability_remote_shards_allnets}
\end{figure}

\subsection{Distributed Inference Overheads}

The P50, P90, and P99 End-to-End (E2E) latency and aggregate compute time overheads, across all sharding strategies, for all models, for serial blocking requests, are shown in Figures~\ref{fig:overheads} and \ref{fig:overheads_rm3}.
A description of each sharding strategy is located in Table~\ref{table:sharding_summary}.
Note that \textit{singular} is the non-distributed case, and \textit{1-shard} is the worse-case with all embedding tables on 1 \remoteshard{}.
P90 and P99 latencies are typical metrics for inference serving because a less accurate fallback recommendation is returned if the inference request is not processed within SLA targets.
P50 is also presented for completeness to show the median case, instead of the mean average, due to long tail latencies discussed in prior work~\cite{fb_rec_hpca}.

\subsection{Latency Overhead}
\label{sec:results_latency}

\subsubsection{Latency overheads can be ameliorated by increasing shards} 
E2E slowdown was incurred for all distributed inference configurations across all models, shown in Figures~\ref{fig:overheads} and \ref{fig:overheads_rm3}.
The parallelization of sparse operators with asynchronous RPC ops was insufficient to overcome the added network latency and software layers.
For \RMone{} and \RMtwo{}, the worst performing configuration, for latency, is unsurprisingly one \remoteshard{}.
This is the impractical worst-case, where all embedding tables are placed on one shard and no work is parallelized.
Encouragingly, at just 2-shards load-balanced, the latency overhead was only 7.3\% at P99 for \RMone{}. 
At 8-shards load-balanced, this overhead fell to just 1\% at P99, and 11\% in the median p50 case.
This demonstrates that it's possible to achieve minimal, practical latency overheads given a simple sharding strategy.
Unexpectedly, the 2-shard NSBP strategy actually had the worst in P99 latency for \RMone{} and nearly the worst for \RMtwo{}. 
Recall in Table~\ref{table:shard_results} that much of the work--pooling factor--is assigned to a single shard, and thus for the P99 case, the 2-shard NSBP configuration acts effectively like a bounding 1-shard configuration.

\subsubsection{Constant overheads eventually dominate}
As the number of shards increase, the work per-shard is reduced and network latency and additional software layers dominate, shown in Figure~\ref{fig:latency_attribution}.
Network latency is measured as the difference between outstanding request time at the main shard, and the total E2E time at the \remoteshard{}.
This time includes in-kernel packet processing and forwarding time.
For all distributed inference configurations, network latency was greater than operator latency.
Distributed inference will always \emph{hurt} the latency of these models.
Put another way, if the sparse operators produced enough work on average, then the model would be amenable to distributed inference.
And given sufficient sparse operator work, latency could be improved.
This provides a multi-discipline opportunity for the system architect, model architect, and feature engineer to collaborate on balancing model resource consumption, performance, and accuracy.

\subsubsection{Sharding impact depends on model architecture}
For \RMthree{}, the number of shards didn't have a strong impact.
\RMthree{} is dominated by a single large table, shown in Figure~\ref{fig:emb_tbl_size}, which is split amongst the shards.
Even at 8-shards, only 2 shards are accessed per inference request--one shard containing the smaller tables and one shard containing the entry for the sharded, largest table, emphasized in Figure~\ref{fig:rm3_per_shard}.

\subsubsection{In-depth latency layer attribution}
Figure~\ref{fig:latency_attribution} shows a breakdown of P50, or median, latency across the layers traced.
Figure~\ref{fig:latency_attribution_e2e} shows that only the embedded portion of the workload is significantly affected across sharding strategies as expected, because this is the portion of the workload being offloaded to \remoteshard{}s.
The \textit{Dense Ops} are all ML operators that are not embedding table lookup and pooling ops; the \textit{Embedded Portion} is all embedding table lookup and pooling operators.
For \textit{singular} this is the ops themselves, whereas for the distributed inference configurations, this is time spent waiting for a response from a \remoteshard{} in Figure~\ref{fig:latency_attribution_e2e}.
\textit{RPC Serde} refers to all serialization and deserialization request times, while \textit{RPC Service} is any other time strictly not spent in a Caffe2 net or serialization/deserialization.
Finally, \textit{Net Overhead} refers to any time in the net not spent executing operators, e.g. scheduling of asynchronous ops.
For distributed inference configurations, the effects on embedded portion latency represent the overhead shown in Figures~\ref{fig:overheads} and \ref{fig:overheads_rm3}.
In the \RMone{} singular configuration, the embedded portion represents only $\sim10\%$ of latency, while for a single-shard, it's 32\%.
The best distributed inference case, for 8-shards load-balanced, it represents 15.6\% of total latency.
\changed{Comparatively, the embedded portion of \RMthree{} does not significantly change as shards are increased because only the large dominating table is further partitioned.
Changes to latency and compute for \RMthree{} are attributed to cache effects and network variability of communicating with more server nodes.}
Figure~\ref{fig:latency_attribution_sls} further attributes latency within that embedded portion--each bar stack represents the embedded portions in Figure~\ref{fig:latency_attribution_e2e}.
For P99, the embedded portion is less significant, and the dense operators and RPC deserialization on the main shard begin to dominate due to very large inference request sizes.
This is why the P99 latency overheads are more favorable than P50.

\subsection{Compute Overhead}
Understanding compute overheads is vital to minimize additional resource requirements at the data-center scale.
Sharding strategy is one method for the system designer to balance latency constraints, discussed in the previous section, and compute overheads which impact resource requirements.

\subsubsection{Increased compute is a trade-off for reduced latency overheads}
The high compute overhead is the trade-off of a flexible, easily deployed system.
As stated in the previous section, increasing \remoteshard{}s can reduce the latency overhead of distributed inference.
However, compute overhead is also increased, because each shard invokes a full \fbthrift{} service.
Figure~\ref{fig:compute_attribution_breakdown} shows that for all models, distributed inference \emph{always} increases compute due to the additional RPC ops required.
More importantly, Figure~\ref{fig:compute_attribution_breakdown} demonstrates that compute \emph{overhead} is proportional to the number of RPC ops.
The NSBP strategy observes the least compute overhead because it executes the least RPC ops.
Recall that the NSBP strategy restricts each shard to not mix embedding tables from different nets, and as such each shard is invoked only once per inference.
Given multiple nets, each net is less parallelized.
Comparatively, the other sharding strategies, which may parallelize each net more, will invoke more RPC ops and lead to increased overall compute overhead.

\subsubsection{Compute overhead impacts data-center resources}
Increased compute overhead is especially problematic when it is incurred on the main shard, because it increases \emph{compute-driven} replication and resource requirements to handle the same QPS.
This occurs when the compute needed to issue RPC ops, on the main shard, dominates the compute saved by offloading the embedded portion.
This is also more likely to occur with model architectures with many, large embedding tables and low pooling factors.
The results provide an impetus to investigate these inflection points, which should be inputs to future automatic sharding methodologies and are dependent on model attributes and software infrastructure.
This is further discussed in Section~\ref{sec:data_center}.

\begin{figure*}[t]
	\begin{subfigure}{.49\textwidth}
		\centering
		\includegraphics[width=\linewidth]{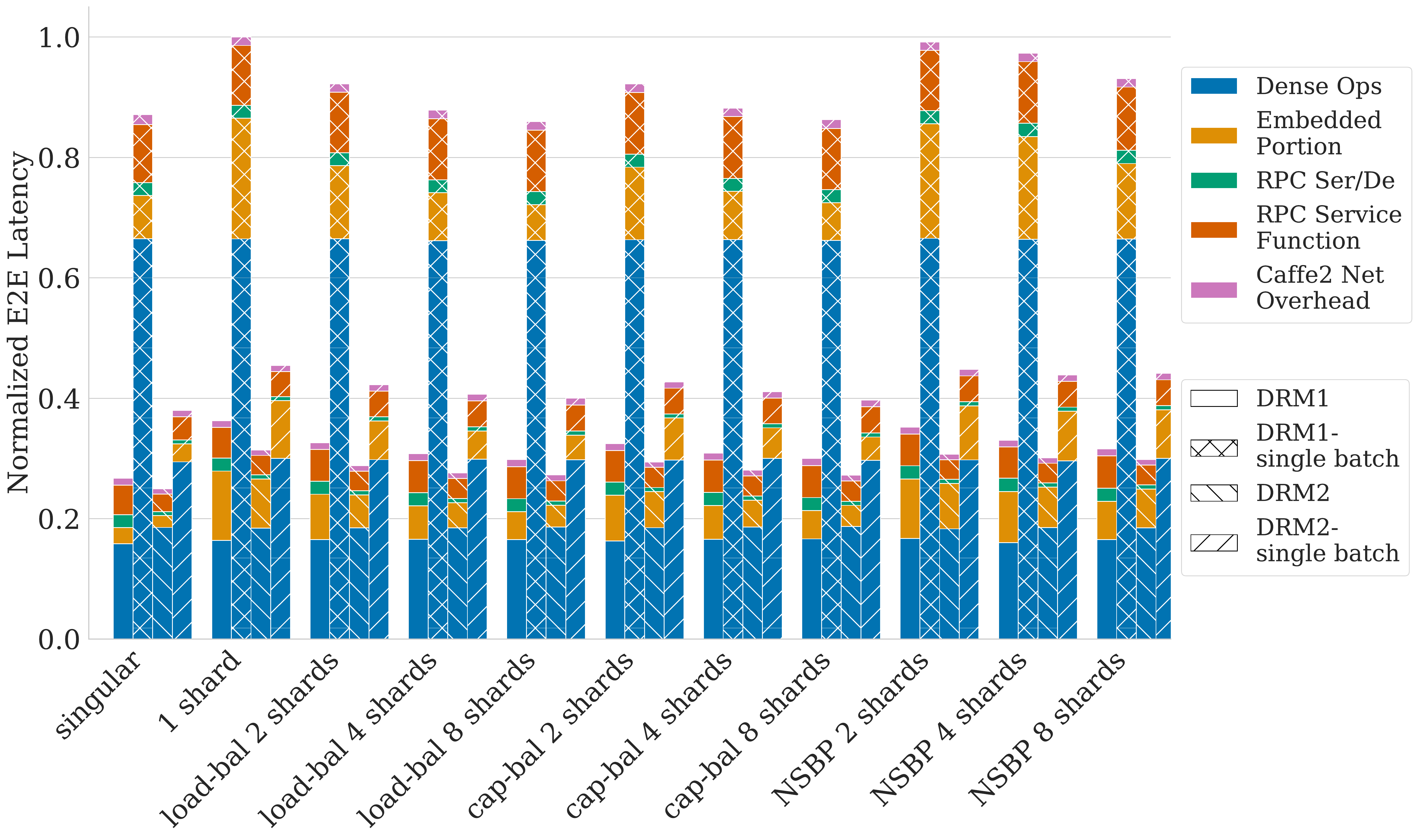}
		\caption{E2E Latency Stack.}
		\label{fig:batch_disabled_lat_e2e}
	\end{subfigure}
	\begin{subfigure}{.49\textwidth}
		\centering
		\includegraphics[width=\linewidth]{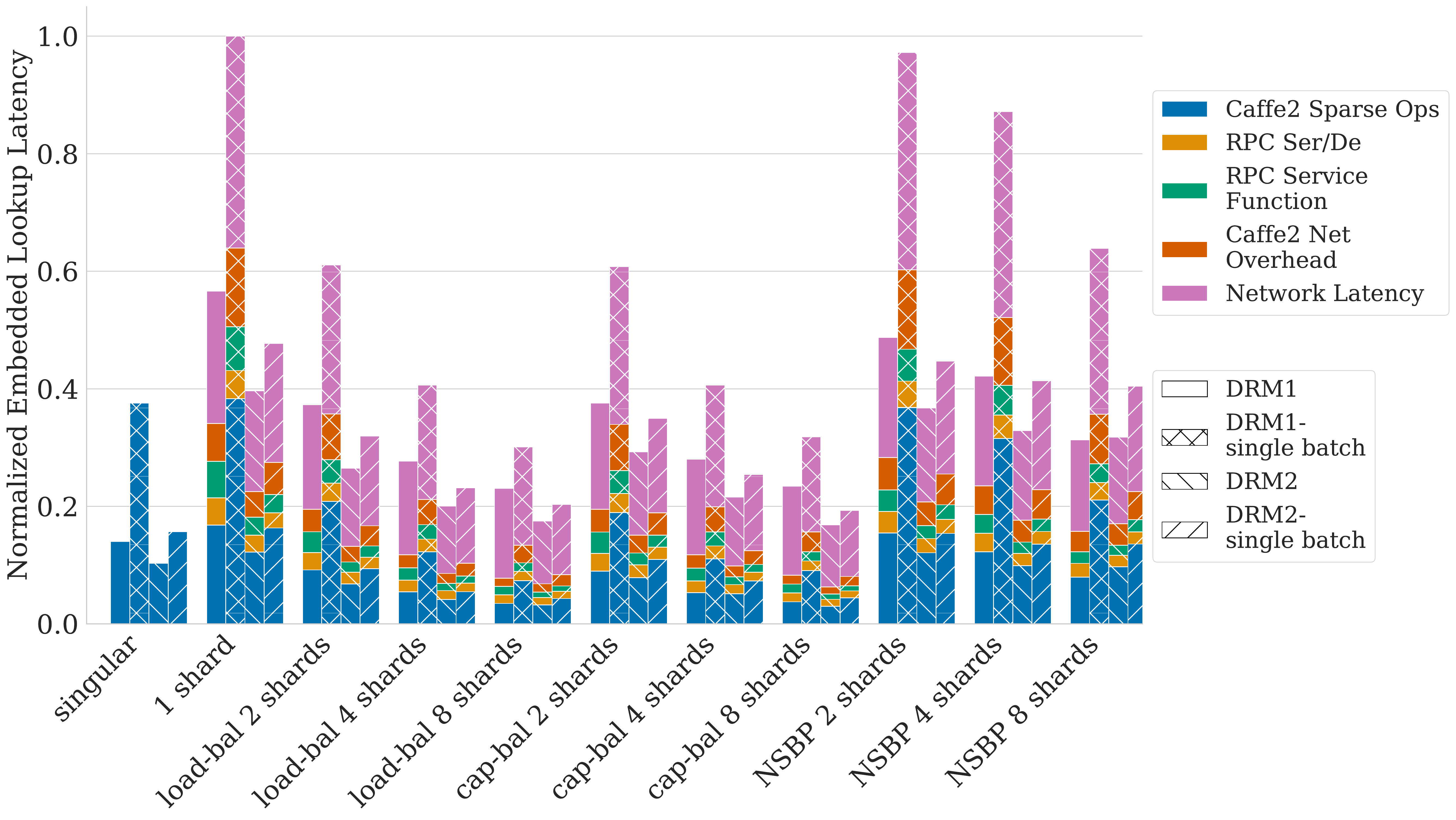}
		\caption{Embedded Portion Latency Stack, for the bounding shard.}
		\label{fig:batch_disabled_lat_emb}
	\end{subfigure}%
	\caption{\changed{\RMone{} \& \RMtwo{} P50 Latency Stacks for default- and single-batch configurations across sharding strategy, normalized to the highest configuration to contrast models. Distributed inference can improve latency given enough work to parallelize in the sparse operators. \RMone{}’s larger requests result in more batches compared to \RMtwo{}.}}
	\label{fig:batch_disabled_lat}
\end{figure*}

\begin{figure}[t]
	\centering
	\includegraphics[width=\linewidth]{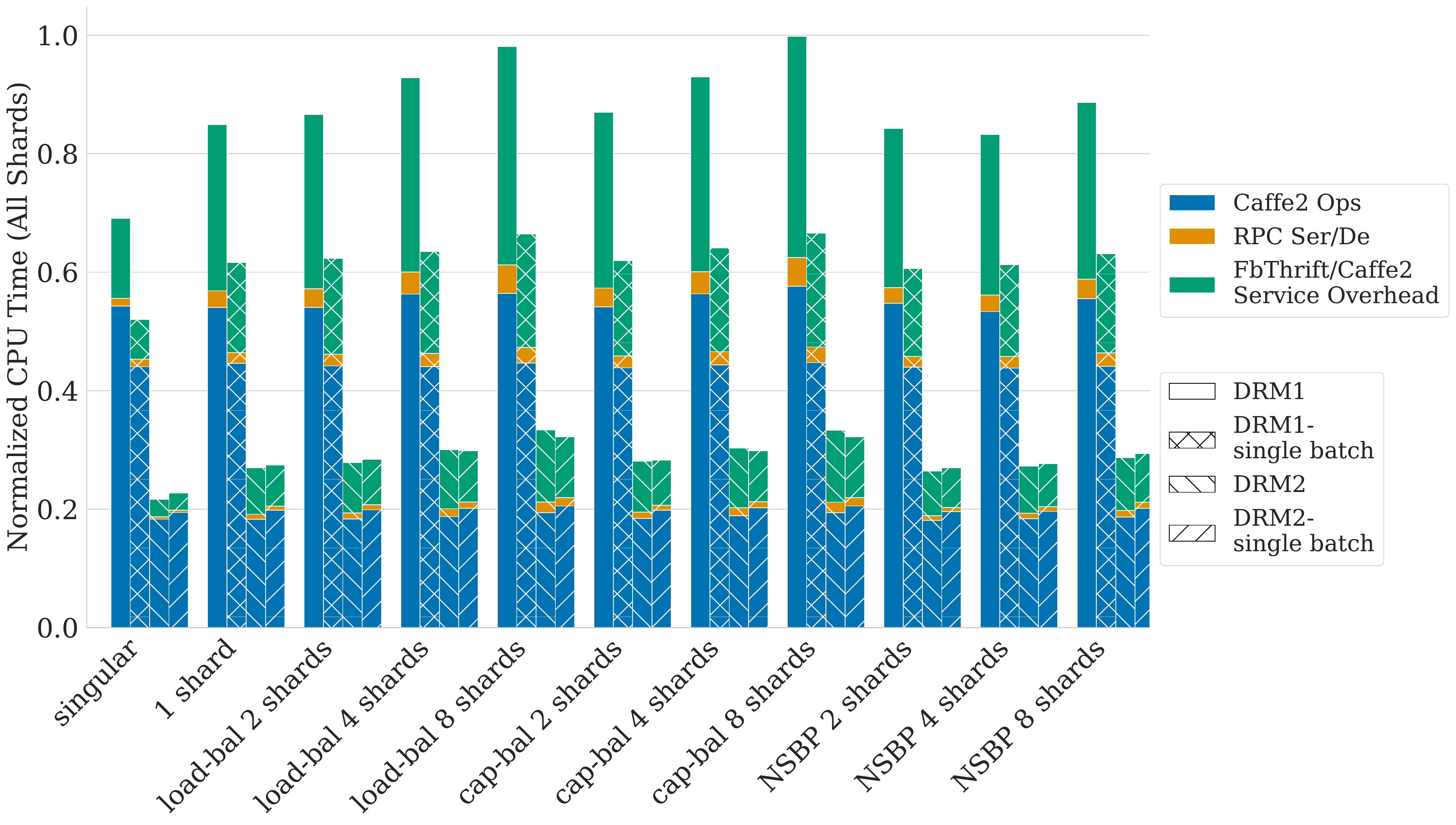}
	\caption{\changed{\RMone{} \& \RMtwo{} P50 CPU Time Stack for default- and single-batch configurations. \RMone{} has more batches than \RMtwo{}, resulting in higher compute overheads.}}
	\label{fig:batch_disabled_comp}
\end{figure}

\subsection{Sharding Strategy Effects}
A trade-off between increased latency or increased compute, as a result of shard count, was established in the previous section.
Shard count is a straightforward knob for system designers to balance compute overheads with latency constraints.
The sharding strategy, discussed in this section, provides another knob for designers, but the effects on latency and compute are more nuanced.

\subsubsection{NSBP is the most scalable strategy}
Latency overheads did not show a strong difference between the load-balanced and capacity-balanced configurations.
However, the net-specific bin-packed strategy deviated by being less impacted by additional sharding.
The most important takeaway from this observation is that NSBP is the most \emph{scalable} strategy evaluated, because it invokes less RPC ops.

NSBP presented the most \emph{unbalanced} per-shard latencies.
Recall in this strategy, embedding tables are first grouped by net, and then assigned to shards based on both net and size.
Tables from separate nets are never assigned to the same shard.
Figure~\ref{fig:latency_variability_remote_shards_pernet} more clearly demonstrates this via the \textit{per-net} operator latencies for the load-balanced and NSBP strategies.
Shards 1 and 2 comprise the first net which performs the most work but has the smallest table sizes, shown in Figure~\ref{fig:latency_variability_remote_shards_rm1_nsbp} and Table~\ref{table:shard_results}.
For NSBP, this had a negative effect on latency since less work is parallelized, however compute overhead is less impacted precisely due to less parallelization, ergo less scheduling and service overheads.
Note that Net1 and Net2 are executed sequentially, so their combined effect on E2E latency is additive.
The benefit of the NSBP strategy, in terms of resource utilization, are further discussed in Section~\ref{sec:data_center}.

\subsubsection{Little overhead difference exists between load-balanced and capacity-balanced strategies}
Figure~\ref{fig:latency_variability_remote_shards_allnets} shows the per-shard operator latencies for all 8-shard configurations with \RMone{}, across all requests; \RMtwo{} shows similar trends and is omitted for brevity.
Recall from Section~\ref{sec:dist_inf_sharding} that load-balanced was expected to have lower latency overheads, by removing \emph{any one shard} from being the bounding critical path.
However, per-shard operator latencies for both strategies are insignificant compared to the E2E latency.
Furthermore, there isn't a significant difference in latencies between load- and capacity-balanced strategies, as was suggested by the estimated pooling factors in Table~\ref{table:shard_results}.
The pooling factors are too small at this scale to show appreciable effect.
For \RMone{} and \RMtwo{}, between load-balanced and capacity-balanced shard strategies, the largest impact comes from increasing the number of shards.

\subsection{Model Variety}
Model attributes also affect distributed inference performance.
The number of nets, number of embedding tables, size distribution, and respective pooling factor are the model attributes most relevant attributes, and are described in Section~\ref{sec:model_test_description} for the evaluated models.
\RMthree{} has less total compute attributed to sparse operators and is dominated by a single large embedding table, compared to the long tail embedding tables in \RMone{} and \RMtwo{}.
Latency for \RMone{} and \RMtwo{} benefit from increased sharding, but \RMthree{} sees no such benefit shown in Figures~\ref{fig:overheads_rm3} and \ref{fig:latency_attribution}.

\subsubsection{Effects of sharding depend on model architecture}
Figure~\ref{fig:rm3_specific} singles out the per-shard operator latency and embedd portion breakdown of \RMthree{} to show additional sharding does not improve latency.
The primary causes are twofold.
(1) Additional sharding only partitions the one capacity-dominating embedding table, which does not parallelize any significant compute.
Figure~\ref{fig:rm3_per_shard} shows shard 1 performing the majority of compute as it contains all embedding tables except for the largest, which is partitioned across the rest of the shards.
(2) Even if the smaller tables were partitioned, the relatively low compute would still preclude practical latency improvements because network latency dominates E2E latency overheads too greatly.
Thus we conclude models with a long tail of embedding tables and higher pooling factors, like \RMone{} and \RMtwo{}, are required to benefit from sharding.

\subsection{Batching Effects}
\label{sec:batch_size}
Batch-sizing for inference splits a request into parallel tasks and is a careful balance between throughput and latency in order to meet SLA and QPS targets.
To show its interaction with distributed inference, we set the batch size artificially large to perform one-batch per-request.
Smaller batch sizes increase task-level parallelism per-request and can reduce latency, but consequently increase task-level overheads and reduce data-parallelism, which can reduce throughput.
In contrast, larger batches increase sparse operator work and benefit from the parallelization of distributed inference.
Batch-sizing for deep recommendation inference is an on-going research topic~\cite{guptadeeprecsys_isca20}.

\subsubsection{Distributed inference improves latency with sufficiently large batch sizes}
Figure~\ref{fig:batch_disabled_lat} shows distributed inference \textit{can improve latency} in the \RMone{} single-batch case, when using 8-shards capacity- or load-balanced configurations.
This is because the sparse operators perform enough work to sufficiently benefit from parallelization, emphasized in Figure~\ref{fig:batch_disabled_lat_emb}.
\RMtwo{} shows similar trends, but is not as strongly impacted because requests are smaller. 
In this context, larger batches can be viewed as a proxy for embedding tables with larger pooling factor, with the salient characteristic being additional lookup indices sent over RPC and increasing sparse operator work and network requirements.
\RMthree{} isn't shown because its requests are typically small enough for only one batch per request, with default batch sizes.

\subsubsection{Batch sizing can manage distributed inference compute overheads}
Figure~\ref{fig:batch_disabled_comp} emphasizes the \emph{multiplicative} compute overhead--each additional batch issues corresponding RPC ops which increases compute requirements.
For example, because NSBP for \RMone{} issues one RPC per shard, its compute overhead increases slower than load-balanced as shards are added with the default-batch size.
With one-batch per-request, the marginal increase in compute from sharding is less severe.
Thus, it’s essential to consider batch-size when exploring distributed inference compute overheads.

\begin{figure}[t]
	\centering
	\includegraphics[width=\linewidth]{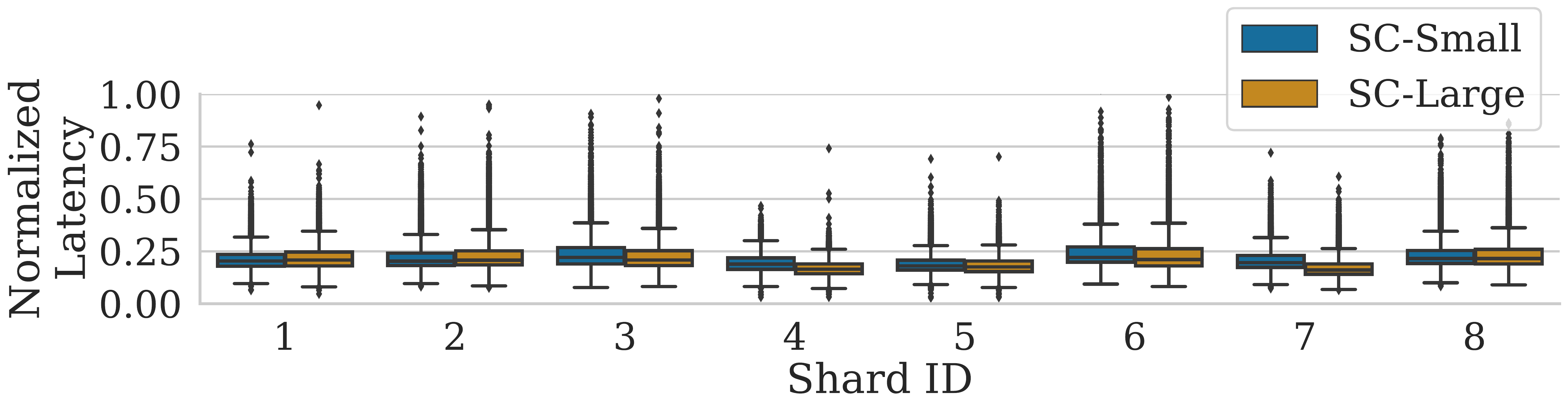}
	\caption{\changed{\RMone{} per-shard operator latencies, by server platform. No significant latency overheads are incurred despite platform differences.}}
	\label{fig:latency_variability_8shards_rm1}
\end{figure}

\begin{figure*}[t]
	\centering
	\includegraphics[width=\linewidth]{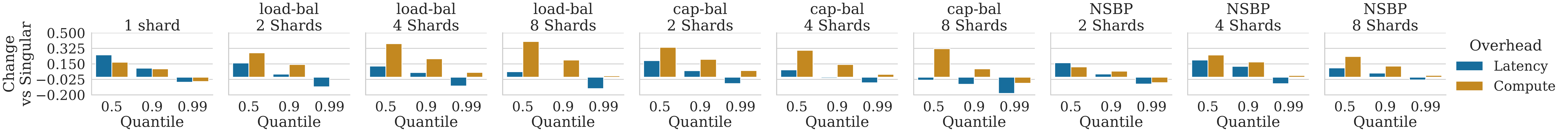}
	\caption{Compute and Latency overheads for \RMone{} at 25QPS. P99 latencies improve over singular for every sharding strategy, including 1-shard.}
	\label{fig:overheads_RM1_25qps}
\end{figure*}

\section{Impacts in Data-Center Environments}
\label{sec:data_center}
In this section, we discuss the implications of distributed inference, for deep recommender systems, in the data-center.
Analyses of latency and compute overheads are important to understanding the impact of distributed inference on SLA targets and additional compute resources.
However, the analyses in Section~\ref{sec:results} focused on a simplified scenario to attribute per-request overheads where (1) each request was processed serially and (2) the servers had the same \SCLarge{} hardware configurations which are over-provisioned.
To model a more representative serving environment, we performed two additional experiments on \RMone{}, which is the most compute intensive model.
First, we sent requests at a higher rate of 25 QPS across all sharding strategies and with the same hardware configuration of \SCLarge{}.
Second, we re-ran the load-balanced configuration across \SCSmall{} platforms more typical for web serving to compare to \SCLarge{}, with requests again sent serially.
We place our results in the context of a serving environment where model instances are replicated to handle real-time traffic.
Lastly, we include a discussion of existing compression techniques currently implemented in large, data-center scale deep recommender systems.
Our primary observations are: 

\begin{itemize}
	\item Requests sent at a higher QPS, indicative of a data-center environment, perform \textit{better} in distributed inference at P99 due to improved resource availability.
	\item Between \SCLarge{} and \SCSmall{}, there is not a significant difference in \remoteshard{} per-request latency, providing an opportunity for better efficiency via \remoteshard{} serving with lower power consumption.
	\item Shard replication provides opportunity for improved serving efficiency by allocating resources for the dense and sparse portions of model independently.
	\item Compression is complementary to distributed inference and cannot address model scalability issues itself.
\end{itemize}

\subsection{High QPS Environment} 
\label{sec:data_center_qps}
The request replayer was configured to send requests at 25QPS to the main server for \RMone{}, the most compute-heavy of all evaluated models.
The overhead graph for each \RMone{} configuration, compared to singular, is shown in Figure~\ref{fig:overheads_RM1_25qps}.
All overheads in the 25QPS experiment are \textit{less} than the same configuration when sent serially, which was shown in Figure~\ref{fig:overheads}.
Across nearly all configurations, the P99 latency is improved compared to the singular configuration, shown in Figure~\ref{fig:overheads}.
Furthermore, in some configurations, like 8-shard capacity- and load-balanced, P50 is either \textit{improved} over singular, or has less than $\sim.05\%$ overhead.
Distributed inference has better latency characteristics in higher QPS scenarios given a model with sufficient sparse operator compute.

\subsection{Sparse Shard Platform Efficiency}
To provide an apples-to-apples comparison of adding servers for distributed inference, the same hardware platform was used for the \remoteshard{}s.
However, Figures~\ref{fig:latency_attribution} and \ref{fig:latency_variability_remote_shards_allnets} shows that the \remoteshard{}s have far less compute requirements compared to the main shard, as the sharding in this work is capacity-driven, not compute-driven.
Thus, we run an additional configuration using lighter-weight \SCSmall{} servers with \RMone{}, which again, is the most compute-intensive model.
Figure~\ref{fig:latency_variability_8shards_rm1} shows the per-shard operator latencies are nearly identical, when run on the original heavier-weight \SCLarge{} and the lighter-weight \SCSmall{} servers.
Recall that the \SCLarge{} server has more and faster cores, and $4\times$ memory capacity, resulting in increased energy footprint.
This suggests the opportunity for coarse-grained platform specialization of \remoteshard{}s for increased serving- and energy-efficiency at the cluster-level.

\subsection{Replication in the Data-Center}
In this section, we include a small discussion on how distributed inference can improve shard replication. 
Supporting singular models that have 100s of GBs of memory footprint requires servers that are inefficiently provisioned.
For example, most of the memory capacity will be dedicated to the large embedding tables, but most cores will be dedicated to using the significantly smaller dense parameters.
The majority of compute touches less than 3\% of the model's memory footprint for \RMone{}, \RMtwo{}, and \RMthree{}.
This inefficiency is scaled to support the QPS of millions or billions of users as inference servers are replicated dynamically.
In such a case, the large load incurred by the dense layers, shown in Figure~\ref{fig:op_stack}, will cause the \textit{entire model} to be replicated to additional servers, including all embedding tables.
Distributed inference alleviates this inefficiency by enabling compute resources to be allocated for dense layers and memory resources to be allocated for sparse layers.
In other words, \emph{the memory requirements of replication are reduced}.
Our heuristic to shard the dense and sparse layers of the models into separate components simplifies this allocation. 
Lastly, sharding strategy plays an auxiliary role in replication.
While the decision to shard sparse operators has the strongest impact on reducing memory requirements of replication, Table~\ref{table:shard_results} shows that sharding strategy can further impact replication.
Recall that \RMone{} is comprised of two primary nets, where Net 2 consumes $4.75\times$ as much memory resources, but 6.3\% of the compute resources as Net 1.
The NSBP sharding strategy constrains each shard to hold either Net 1 or Net 2, and accordingly doesn't parallelize work as well compared a net-agnostic sharding strategy.
However, NSBP can further improve resource utilization by grouping the most compute-dense embedding tables together, even at higher shard counts necessary for larger models.
This sharding strategy trade-off, between improved latency and improved resource utilization, will need to be made on a model-by-model basis and encourages further work in sharding automation.

\subsection{Model Compression}
\label{sec:compression}

\begin{table}[t]
	\caption{Effect of Quantization and Pruning on \RMone{}. CPU Times and E2E Latencies are normalized to the respective Uncompressed P50 values.}
	\begin{center}
		\begin{tabularx}{\linewidth}{| >{\hsize=0.55\hsize}Y | >{\hsize=0.2\hsize}X | Y | Y | }
			\hline
			\multicolumn{2}{|c|}{}& Uncompressed &  Quantized and Pruned \\
			\Xhline{4\arrayrulewidth}
			\multicolumn{2}{|l|}{Total Size (GB) } & 194.46 & 35 (5.56$\times$) \\
			\Xhline{4\arrayrulewidth}
			\multirow{3}{\hsize}{CPU Time} 
			& P50 & 1$\times$ & 0.98$\times$  \\
			\cline{2-4}
			& P90 & 3.53$\times$ & 3.47$\times$  \\
			\cline{2-4}
			& P99 & 6.6$\times$ & 6.31$\times$ \\
			\Xhline{4\arrayrulewidth}
			\multirow{3}{\hsize}{E2E Latency} 
			& P50 & 1$\times$ &  .99$\times$ \\
			\cline{2-4}
			& P90 & 2.72$\times$ & 2.75$\times$  \\
			\cline{2-4}
			& P99 & 5.03$\times$ & 4.87$\times$ \\
			\hline
		\end{tabularx}
		\label{table:compression_singular}
	\end{center}
\end{table}

Model compression is the traditional approach to constrain model capacity.
We show the compressed model size for \RMone{} in Table~\ref{table:compression_singular} as a point-of-reference.
All compression techniques deployed on current data-center models were used to generate the compressed model, which utilize quantization and pruning\cite{deep_compression_iclr}.
All tables were row-wise linear quantized to at least 8-bits, and sufficiently large tables were quantized to 4-bits.
Tables were manually pruned as specified by the model architect based on a threshold magnitude or training update frequency.
Quantization and pruning options were chosen to preserve accuracy and latency.
We note that latency and compute are marginally \emph{improved} with compression, but because this was a focus of this work, we leave this analysis to future work.
We speculate the cause is improved memory locality, relatively. 
More relevantly, Table~\ref{table:compression_singular} shows the compressed model is 5.56$\times$ smaller.
While significant, even with these savings, large models will still not be able to fit on one, two, or even four commodity servers configured with $\sim$50GB of usable DRAM.
Thus, compression alone is insufficient to enable emerging large \dlrm{}s.

\section{Related Work}
\label{sec:related_work}

Recent work on inference for deep learning recommendation models focused on smaller models that fit on a single machine~\cite{guptadeeprecsys_isca20, fb_dlrm, fb_rec_hpca, fb_recnmp, fb_bandana, hwang2020centaur}.
Characteristics like compute density, embedding table memory access patterns, and batch-size effects were explored in the context of CPU-architecture and accelerators.
However, these inference-focused works did not explore the trends of scaling models to the terabyte-scale.

Separate works have described the challenges of scaling \emph{training} for large models, but they either don't address \dlrm{}s or don't explore the inference serving environment that is the focus of our paper~\cite{gshard, gpipe, mesh_tensorflow, fb_scaleupout, baidu_dist_inf, deepretrieval}.  
Google's data-center scale ML serving infrastructure, TensorFlow-Serving, encompasses model loading, versioning, RPC APIs, and inference batching~\cite{tf_serving}.
Support for data-parallel distributed inference exists as part of distributed TensorFlow~\cite{tensorflow}, but the distributed schemes described in this work are not currently supported out-of-the-box.
The remote operators used are similar to the ones described in this work, but a distributed serving paradigm has only briefly been mentioned in publication~\cite{deepretrieval}.
More recent works have discussed infrastructure for model-parallelism in TensorFlow--and in turn huge models--but have focused on language models and training which don't have the variety and number of features shown in Figure~\ref{fig:emb_tbl_size} and emphasize training throughput compared to inference's latency-bounded throughput~\cite{gshard, gpipe, mesh_tensorflow, pipedream, flexflow}.

Gshard provides a sharding abstraction that strongly relies on manual user annotation for tensor placement, and it does not automatically shard based on dynamic inputs as is explored in this work~\cite{gshard}.
Furthermore, Baidu's recent work includes multi-GPU, fast SSDs, and pipelined stages specifically to support large embedding tables for recommendation~\cite{baidu_aibox, baidu_dist_inf}.
In that work, compared to the baseline MPI solution, their GPU+SSD system attains higher throughput during training and requires less physical nodes.
However, such a system is more complex to deploy and the effects on end-to-end latency and strict SLA targets is unexplored in the serving environment.
In both cases, the evaluated use-case is parallelism for training where model parallelism is mostly static, compared to the dynamic environment of an inference serving, where model components can be replicated to meet influxes of QPS.
While \textit{training} massive \dlrm{}s is also an incredible and important challenge--and is related to inference serving--this work focuses on the data-center implications of \textit{serving} those models.
The sharding strategies discussed in this work are heuristically tailored to the unique model architecture of \dlrm{}s and driven by capacity and latency constraints, not training throughput.

Cooperative, hybrid approaches--that split the model between mobile systems and data-center--have also been explored for improved performance and efficiency~\cite{neurosurgeon}.
However, such a scheme is problematic in the scope of recommendation where the model is already capacity-bound, so execution on a constrained mobile system is challenging.
Privacy requirements, energy concerns for the user, and online models further impede the placement of model partitions on user end devices.

Model compression is discussed in Section~\ref{sec:compression}, which concluded current quantization and pruning techniques do not solely resolve the challenge of serving terabyte-scale \dlrm{}s.
The effectiveness of traditional compression techniques is specific to the neural network model, with CNNs in particular shown to compress well~\cite{deep_compression_iclr}. 
Techniques specific to large embedding tables have traditionally targeted the intuitive characteristics of pre-trained word embeddings, not the sparse user- and content-features of \dlrm{}s~\cite{word2vec, glove, compressingwordemb, allbutthetop, wordembdims, wordembred, tt_embedding}.
Other lossless in-memory or in-cache compression techniques assume data-sets that exhibit lower entropy and more regularity than observed in embedding tables~\cite{farmemory, bdi, fpc}.
Even quantized to 8- or 4- bits, embedding tables value distributions do not significantly benefit from further lossless entropy encoding.
Techniques to reduce the size of recommendation embedding tables warrants further exploration and is an on-going area of research\cite{fb_compositional_embeddings}.

Lastly, the distributed scheme described in this work, that shards only the embedding tables, can be compared to traditional scale-out, in-memory distributed databases or hash tables that serve large magnitudes of tabular data~\cite{hadoop, bigtable}.
However, these are overly complex due to their mutability and scale requirements, compared to a trained model snapshot.
Even within training, the varying levels of asynchrony of parameter servers also does not lend itself to existing distributed database solutions.
Given a key motivation of distributed inference is rapid and flexible deployment, integration with database solutions was not considered.

\section{Industry and Academic Relevance}
\label{sec:academic_relevance}
The challenge of large deep learning recommendation models provides a strong opportunity for academics to tackle problems of the data-center, the impact of which is significant due to the ubiquity of \dlrm{}s.
A core contribution of this work is that the models and infrastructure are highly representative of real inference serving.
The software infrastructure in particular matches the data center environment.
Despite the seemingly industry-oriented inclination of this workload, academic-friendly methodologies also exist that are less resource intensive.
DLRM is an open-source recommendation model that has been used in many recent works~\cite{fb_dlrm, fb_rec_hpca, fb_recnmp, fb_bandana, guptadeeprecsys_isca20}.
While it currently doesn't support \textit{distributed} inference, the opportunity to extend this functionality is straightforward with the RPC operators described in Section~\ref{sec:dist_inf}.
Further research opportunities are achievable by way of trace-driven experimentation.
For example, Bandana used embedding table access traces--which can be collected offline--to reduce effective DRAM requirements~\cite{fb_bandana}.
Because embedding table behavior is the dominating design factor in large models, explorations table placement and frequency-based caching are also valuable directions enabled with trace-based analyses.

\section{Conclusion and Future Work}
\label{sec:conclusion}
The rate and scale of model growth for recommendation motivates new serving paradigms to accommodate enormous sparse embedding tables.
This emerging domain provides rich opportunity for system architects to have impact.  
Distributed inference in particular provides a solution that is easily deployed by existing infrastructure.
Understanding the performance trade-offs and overhead sources is paramount to efficient and scalable recommendation.

This is the first work to describe and characterize capacity-driven distributed inference for deep learning recommender systems, with a cross-layer distributed tracing methodology.
In this work, we described and identified distributed inference as a practical, deployable solution on existing data-center infrastructure.
Even with the naive sharding strategies we evaluated, only a 1\% increase in P99 latency was experienced at 8-shards for the most compute-intensive model.
The trace-based characterization identified sharding strategy, sparse feature hyperparameters like pooling factor, and request size as having substantial impact on latency and compute overheads.
In particular, we found that organizing shards by net had a stronger impact on both latency and compute overheads, compared to only considering embedding tables in isolation.
These results varied depending on the model and warrants a workflow that dynamically profiles models.
We also showed that--given enough sparse operator work--latency can be \textit{improved} over the non-distributed model, especially when batches are sufficiently large and in a high QPS serving environment.
Finally, potentials for improved data-center scale serving efficiency were identified, by decoupling dense and sparse operator resources in distributed inference.

Future work is needed to automate model sharding to target data-center resource efficiency and per-model SLA and QPS requirements.
The various design trade-offs identified in this work place a large burden on the feature, model, and system designers to manually optimize the distributed deep learning recommendation model. 
Moreover, the design space should be expanded to include additional system-level solutions such as paging-from-disk, and to include additional large models, such as recent terabyte-scale NLP models and graph-based recommender systems~\cite{gshard, gcn_rec}.

\section*{Acknowledgments}

This work was completed in a collaboration between Drexel University, Tufts University, and Facebook.
The majority of work was completed while some authors were on internship and sabbatical at Facebook.
We would like to thank our colleagues in Facebook AI for their feedback, key insights, and myriad of discussions on recommender systems and distributed infrastructures and AI.
The authors would like to specifically thank Udit Gupta, Liu Ke, Vikram Saraph, Mark Jeffrey, Caroline Trippel, Brandon Reagen, Michael Bevilacqua-Linn, Tristan Rice, Xuan Zhang, Hsien-Hsin Sean Lee, and Baris Taskin for their invaluable mentorship, guidance and support, without whom this work would not be possible.

\pagebreak
\bibliographystyle{IEEEtran}
\bibliography{references}

\end{document}